\documentclass[useAMS,usenatbib,usegraphicx]{mn2e}
\usepackage{color}
\definecolor{AliceBlue}{rgb}{0.94,0.97,1.00}
\definecolor{AntiqueWhite1}{rgb}{1.00,0.94,0.86}
\definecolor{AntiqueWhite2}{rgb}{0.93,0.87,0.80}
\definecolor{AntiqueWhite3}{rgb}{0.80,0.75,0.69}
\definecolor{AntiqueWhite4}{rgb}{0.55,0.51,0.47}
\definecolor{AntiqueWhite}{rgb}{0.98,0.92,0.84}
\definecolor{BlanchedAlmond}{rgb}{1.00,0.92,0.80}
\definecolor{BlueViolet}{rgb}{0.54,0.17,0.89}
\definecolor{CadetBlue1}{rgb}{0.60,0.96,1.00}
\definecolor{CadetBlue2}{rgb}{0.56,0.90,0.93}
\definecolor{CadetBlue3}{rgb}{0.48,0.77,0.80}
\definecolor{CadetBlue4}{rgb}{0.33,0.53,0.55}
\definecolor{CadetBlue}{rgb}{0.37,0.62,0.63}
\definecolor{CornflowerBlue}{rgb}{0.39,0.58,0.93}
\definecolor{DarkBlue}{rgb}{0.00,0.00,0.55}
\definecolor{DarkCyan}{rgb}{0.00,0.55,0.55}
\definecolor{DarkGoldenrod1}{rgb}{1.00,0.73,0.06}
\definecolor{DarkGoldenrod2}{rgb}{0.93,0.68,0.05}
\definecolor{DarkGoldenrod3}{rgb}{0.80,0.58,0.05}
\definecolor{DarkGoldenrod4}{rgb}{0.55,0.40,0.03}
\definecolor{DarkGoldenrod}{rgb}{0.72,0.53,0.04}
\definecolor{DarkGray}{rgb}{0.66,0.66,0.66}
\definecolor{DarkGreen}{rgb}{0.00,0.39,0.00}
\definecolor{DarkGrey}{rgb}{0.66,0.66,0.66}
\definecolor{DarkKhaki}{rgb}{0.74,0.72,0.42}
\definecolor{DarkMagenta}{rgb}{0.55,0.00,0.55}
\definecolor{DarkOliveGreen1}{rgb}{0.79,1.00,0.44}
\definecolor{DarkOliveGreen2}{rgb}{0.74,0.93,0.41}
\definecolor{DarkOliveGreen3}{rgb}{0.64,0.80,0.35}
\definecolor{DarkOliveGreen4}{rgb}{0.43,0.55,0.24}
\definecolor{DarkOliveGreen}{rgb}{0.33,0.42,0.18}
\definecolor{DarkOrange1}{rgb}{1.00,0.50,0.00}
\definecolor{DarkOrange2}{rgb}{0.93,0.46,0.00}
\definecolor{DarkOrange3}{rgb}{0.80,0.40,0.00}
\definecolor{DarkOrange4}{rgb}{0.55,0.27,0.00}
\definecolor{DarkOrange}{rgb}{1.00,0.55,0.00}
\definecolor{DarkOrchid1}{rgb}{0.75,0.24,1.00}
\definecolor{DarkOrchid2}{rgb}{0.70,0.23,0.93}
\definecolor{DarkOrchid3}{rgb}{0.60,0.20,0.80}
\definecolor{DarkOrchid4}{rgb}{0.41,0.13,0.55}
\definecolor{DarkOrchid}{rgb}{0.60,0.20,0.80}
\definecolor{DarkRed}{rgb}{0.55,0.00,0.00}
\definecolor{DarkSalmon}{rgb}{0.91,0.59,0.48}
\definecolor{DarkSeaGreen1}{rgb}{0.76,1.00,0.76}
\definecolor{DarkSeaGreen2}{rgb}{0.71,0.93,0.71}
\definecolor{DarkSeaGreen3}{rgb}{0.61,0.80,0.61}
\definecolor{DarkSeaGreen4}{rgb}{0.41,0.55,0.41}
\definecolor{DarkSeaGreen}{rgb}{0.56,0.74,0.56}
\definecolor{DarkSlateBlue}{rgb}{0.28,0.24,0.55}
\definecolor{DarkSlateGray1}{rgb}{0.59,1.00,1.00}
\definecolor{DarkSlateGray2}{rgb}{0.55,0.93,0.93}
\definecolor{DarkSlateGray3}{rgb}{0.47,0.80,0.80}
\definecolor{DarkSlateGray4}{rgb}{0.32,0.55,0.55}
\definecolor{DarkSlateGray}{rgb}{0.18,0.31,0.31}
\definecolor{DarkSlateGrey}{rgb}{0.18,0.31,0.31}
\definecolor{DarkTurquoise}{rgb}{0.00,0.81,0.82}
\definecolor{DarkViolet}{rgb}{0.58,0.00,0.83}
\definecolor{DeepPink1}{rgb}{1.00,0.08,0.58}
\definecolor{DeepPink2}{rgb}{0.93,0.07,0.54}
\definecolor{DeepPink3}{rgb}{0.80,0.06,0.46}
\definecolor{DeepPink4}{rgb}{0.55,0.04,0.31}
\definecolor{DeepPink}{rgb}{1.00,0.08,0.58}
\definecolor{DeepSkyBlue1}{rgb}{0.00,0.75,1.00}
\definecolor{DeepSkyBlue2}{rgb}{0.00,0.70,0.93}
\definecolor{DeepSkyBlue3}{rgb}{0.00,0.60,0.80}
\definecolor{DeepSkyBlue4}{rgb}{0.00,0.41,0.55}
\definecolor{DeepSkyBlue}{rgb}{0.00,0.75,1.00}
\definecolor{DimGray}{rgb}{0.41,0.41,0.41}
\definecolor{DimGrey}{rgb}{0.41,0.41,0.41}
\definecolor{DodgerBlue1}{rgb}{0.12,0.56,1.00}
\definecolor{DodgerBlue2}{rgb}{0.11,0.53,0.93}
\definecolor{DodgerBlue3}{rgb}{0.09,0.45,0.80}
\definecolor{DodgerBlue4}{rgb}{0.06,0.31,0.55}
\definecolor{DodgerBlue}{rgb}{0.12,0.56,1.00}
\definecolor{FloralWhite}{rgb}{1.00,0.98,0.94}
\definecolor{ForestGreen}{rgb}{0.13,0.55,0.13}
\definecolor{GhostWhite}{rgb}{0.97,0.97,1.00}
\definecolor{GreenYellow}{rgb}{0.68,1.00,0.18}
\definecolor{HotPink1}{rgb}{1.00,0.43,0.71}
\definecolor{HotPink2}{rgb}{0.93,0.42,0.65}
\definecolor{HotPink3}{rgb}{0.80,0.38,0.56}
\definecolor{HotPink4}{rgb}{0.55,0.23,0.38}
\definecolor{HotPink}{rgb}{1.00,0.41,0.71}
\definecolor{IndianRed1}{rgb}{1.00,0.42,0.42}
\definecolor{IndianRed2}{rgb}{0.93,0.39,0.39}
\definecolor{IndianRed3}{rgb}{0.80,0.33,0.33}
\definecolor{IndianRed4}{rgb}{0.55,0.23,0.23}
\definecolor{IndianRed}{rgb}{0.80,0.36,0.36}
\definecolor{LavenderBlush1}{rgb}{1.00,0.94,0.96}
\definecolor{LavenderBlush2}{rgb}{0.93,0.88,0.90}
\definecolor{LavenderBlush3}{rgb}{0.80,0.76,0.77}
\definecolor{LavenderBlush4}{rgb}{0.55,0.51,0.53}
\definecolor{LavenderBlush}{rgb}{1.00,0.94,0.96}
\definecolor{LawnGreen}{rgb}{0.49,0.99,0.00}
\definecolor{LemonChiffon1}{rgb}{1.00,0.98,0.80}
\definecolor{LemonChiffon2}{rgb}{0.93,0.91,0.75}
\definecolor{LemonChiffon3}{rgb}{0.80,0.79,0.65}
\definecolor{LemonChiffon4}{rgb}{0.55,0.54,0.44}
\definecolor{LemonChiffon}{rgb}{1.00,0.98,0.80}
\definecolor{LightBlue1}{rgb}{0.75,0.94,1.00}
\definecolor{LightBlue2}{rgb}{0.70,0.87,0.93}
\definecolor{LightBlue3}{rgb}{0.60,0.75,0.80}
\definecolor{LightBlue4}{rgb}{0.41,0.51,0.55}
\definecolor{LightBlue}{rgb}{0.68,0.85,0.90}
\definecolor{LightCoral}{rgb}{0.94,0.50,0.50}
\definecolor{LightCyan1}{rgb}{0.88,1.00,1.00}
\definecolor{LightCyan2}{rgb}{0.82,0.93,0.93}
\definecolor{LightCyan3}{rgb}{0.71,0.80,0.80}
\definecolor{LightCyan4}{rgb}{0.48,0.55,0.55}
\definecolor{LightCyan}{rgb}{0.88,1.00,1.00}
\definecolor{LightGoldenrod1}{rgb}{1.00,0.93,0.55}
\definecolor{LightGoldenrod2}{rgb}{0.93,0.86,0.51}
\definecolor{LightGoldenrod3}{rgb}{0.80,0.75,0.44}
\definecolor{LightGoldenrod4}{rgb}{0.55,0.51,0.30}
\definecolor{LightGoldenrodYellow}{rgb}{0.98,0.98,0.82}
\definecolor{LightGoldenrod}{rgb}{0.93,0.87,0.51}
\definecolor{LightGray}{rgb}{0.83,0.83,0.83}
\definecolor{LightGreen}{rgb}{0.56,0.93,0.56}
\definecolor{LightGrey}{rgb}{0.83,0.83,0.83}
\definecolor{LightPink1}{rgb}{1.00,0.68,0.73}
\definecolor{LightPink2}{rgb}{0.93,0.64,0.68}
\definecolor{LightPink3}{rgb}{0.80,0.55,0.58}
\definecolor{LightPink4}{rgb}{0.55,0.37,0.40}
\definecolor{LightPink}{rgb}{1.00,0.71,0.76}
\definecolor{LightSalmon1}{rgb}{1.00,0.63,0.48}
\definecolor{LightSalmon2}{rgb}{0.93,0.58,0.45}
\definecolor{LightSalmon3}{rgb}{0.80,0.51,0.38}
\definecolor{LightSalmon4}{rgb}{0.55,0.34,0.26}
\definecolor{LightSalmon}{rgb}{1.00,0.63,0.48}
\definecolor{LightSeaGreen}{rgb}{0.13,0.70,0.67}
\definecolor{LightSkyBlue1}{rgb}{0.69,0.89,1.00}
\definecolor{LightSkyBlue2}{rgb}{0.64,0.83,0.93}
\definecolor{LightSkyBlue3}{rgb}{0.55,0.71,0.80}
\definecolor{LightSkyBlue4}{rgb}{0.38,0.48,0.55}
\definecolor{LightSkyBlue}{rgb}{0.53,0.81,0.98}
\definecolor{LightSlateBlue}{rgb}{0.52,0.44,1.00}
\definecolor{LightSlateGray}{rgb}{0.47,0.53,0.60}
\definecolor{LightSlateGrey}{rgb}{0.47,0.53,0.60}
\definecolor{LightSteelBlue1}{rgb}{0.79,0.88,1.00}
\definecolor{LightSteelBlue2}{rgb}{0.74,0.82,0.93}
\definecolor{LightSteelBlue3}{rgb}{0.64,0.71,0.80}
\definecolor{LightSteelBlue4}{rgb}{0.43,0.48,0.55}
\definecolor{LightSteelBlue}{rgb}{0.69,0.77,0.87}
\definecolor{LightYellow1}{rgb}{1.00,1.00,0.88}
\definecolor{LightYellow2}{rgb}{0.93,0.93,0.82}
\definecolor{LightYellow3}{rgb}{0.80,0.80,0.71}
\definecolor{LightYellow4}{rgb}{0.55,0.55,0.48}
\definecolor{LightYellow}{rgb}{1.00,1.00,0.88}
\definecolor{LimeGreen}{rgb}{0.20,0.80,0.20}
\definecolor{MediumAquamarine}{rgb}{0.40,0.80,0.67}
\definecolor{MediumBlue}{rgb}{0.00,0.00,0.80}
\definecolor{MediumOrchid1}{rgb}{0.88,0.40,1.00}
\definecolor{MediumOrchid2}{rgb}{0.82,0.37,0.93}
\definecolor{MediumOrchid3}{rgb}{0.71,0.32,0.80}
\definecolor{MediumOrchid4}{rgb}{0.48,0.22,0.55}
\definecolor{MediumOrchid}{rgb}{0.73,0.33,0.83}
\definecolor{MediumPurple1}{rgb}{0.67,0.51,1.00}
\definecolor{MediumPurple2}{rgb}{0.62,0.47,0.93}
\definecolor{MediumPurple3}{rgb}{0.54,0.41,0.80}
\definecolor{MediumPurple4}{rgb}{0.36,0.28,0.55}
\definecolor{MediumPurple}{rgb}{0.58,0.44,0.86}
\definecolor{MediumSeaGreen}{rgb}{0.24,0.70,0.44}
\definecolor{MediumSlateBlue}{rgb}{0.48,0.41,0.93}
\definecolor{MediumSpringGreen}{rgb}{0.00,0.98,0.60}
\definecolor{MediumTurquoise}{rgb}{0.28,0.82,0.80}
\definecolor{MediumVioletRed}{rgb}{0.78,0.08,0.52}
\definecolor{MidnightBlue}{rgb}{0.10,0.10,0.44}
\definecolor{MintCream}{rgb}{0.96,1.00,0.98}
\definecolor{MistyRose1}{rgb}{1.00,0.89,0.88}
\definecolor{MistyRose2}{rgb}{0.93,0.84,0.82}
\definecolor{MistyRose3}{rgb}{0.80,0.72,0.71}
\definecolor{MistyRose4}{rgb}{0.55,0.49,0.48}
\definecolor{MistyRose}{rgb}{1.00,0.89,0.88}
\definecolor{NavajoWhite1}{rgb}{1.00,0.87,0.68}
\definecolor{NavajoWhite2}{rgb}{0.93,0.81,0.63}
\definecolor{NavajoWhite3}{rgb}{0.80,0.70,0.55}
\definecolor{NavajoWhite4}{rgb}{0.55,0.47,0.37}
\definecolor{NavajoWhite}{rgb}{1.00,0.87,0.68}
\definecolor{NavyBlue}{rgb}{0.00,0.00,0.50}
\definecolor{OldLace}{rgb}{0.99,0.96,0.90}
\definecolor{OliveDrab1}{rgb}{0.75,1.00,0.24}
\definecolor{OliveDrab2}{rgb}{0.70,0.93,0.23}
\definecolor{OliveDrab3}{rgb}{0.60,0.80,0.20}
\definecolor{OliveDrab4}{rgb}{0.41,0.55,0.13}
\definecolor{OliveDrab}{rgb}{0.42,0.56,0.14}
\definecolor{OrangeRed1}{rgb}{1.00,0.27,0.00}
\definecolor{OrangeRed2}{rgb}{0.93,0.25,0.00}
\definecolor{OrangeRed3}{rgb}{0.80,0.22,0.00}
\definecolor{OrangeRed4}{rgb}{0.55,0.15,0.00}
\definecolor{OrangeRed}{rgb}{1.00,0.27,0.00}
\definecolor{PaleGoldenrod}{rgb}{0.93,0.91,0.67}
\definecolor{PaleGreen1}{rgb}{0.60,1.00,0.60}
\definecolor{PaleGreen2}{rgb}{0.56,0.93,0.56}
\definecolor{PaleGreen3}{rgb}{0.49,0.80,0.49}
\definecolor{PaleGreen4}{rgb}{0.33,0.55,0.33}
\definecolor{PaleGreen}{rgb}{0.60,0.98,0.60}
\definecolor{PaleTurquoise1}{rgb}{0.73,1.00,1.00}
\definecolor{PaleTurquoise2}{rgb}{0.68,0.93,0.93}
\definecolor{PaleTurquoise3}{rgb}{0.59,0.80,0.80}
\definecolor{PaleTurquoise4}{rgb}{0.40,0.55,0.55}
\definecolor{PaleTurquoise}{rgb}{0.69,0.93,0.93}
\definecolor{PaleVioletRed1}{rgb}{1.00,0.51,0.67}
\definecolor{PaleVioletRed2}{rgb}{0.93,0.47,0.62}
\definecolor{PaleVioletRed3}{rgb}{0.80,0.41,0.54}
\definecolor{PaleVioletRed4}{rgb}{0.55,0.28,0.36}
\definecolor{PaleVioletRed}{rgb}{0.86,0.44,0.58}
\definecolor{PapayaWhip}{rgb}{1.00,0.94,0.84}
\definecolor{PeachPuff1}{rgb}{1.00,0.85,0.73}
\definecolor{PeachPuff2}{rgb}{0.93,0.80,0.68}
\definecolor{PeachPuff3}{rgb}{0.80,0.69,0.58}
\definecolor{PeachPuff4}{rgb}{0.55,0.47,0.40}
\definecolor{PeachPuff}{rgb}{1.00,0.85,0.73}
\definecolor{PowderBlue}{rgb}{0.69,0.88,0.90}
\definecolor{RosyBrown1}{rgb}{1.00,0.76,0.76}
\definecolor{RosyBrown2}{rgb}{0.93,0.71,0.71}
\definecolor{RosyBrown3}{rgb}{0.80,0.61,0.61}
\definecolor{RosyBrown4}{rgb}{0.55,0.41,0.41}
\definecolor{RosyBrown}{rgb}{0.74,0.56,0.56}
\definecolor{RoyalBlue1}{rgb}{0.28,0.46,1.00}
\definecolor{RoyalBlue2}{rgb}{0.26,0.43,0.93}
\definecolor{RoyalBlue3}{rgb}{0.23,0.37,0.80}
\definecolor{RoyalBlue4}{rgb}{0.15,0.25,0.55}
\definecolor{RoyalBlue}{rgb}{0.25,0.41,0.88}
\definecolor{SaddleBrown}{rgb}{0.55,0.27,0.07}
\definecolor{SandyBrown}{rgb}{0.96,0.64,0.38}
\definecolor{SeaGreen1}{rgb}{0.33,1.00,0.62}
\definecolor{SeaGreen2}{rgb}{0.31,0.93,0.58}
\definecolor{SeaGreen3}{rgb}{0.26,0.80,0.50}
\definecolor{SeaGreen4}{rgb}{0.18,0.55,0.34}
\definecolor{SeaGreen}{rgb}{0.18,0.55,0.34}
\definecolor{SkyBlue1}{rgb}{0.53,0.81,1.00}
\definecolor{SkyBlue2}{rgb}{0.49,0.75,0.93}
\definecolor{SkyBlue3}{rgb}{0.42,0.65,0.80}
\definecolor{SkyBlue4}{rgb}{0.29,0.44,0.55}
\definecolor{SkyBlue}{rgb}{0.53,0.81,0.92}
\definecolor{SlateBlue1}{rgb}{0.51,0.44,1.00}
\definecolor{SlateBlue2}{rgb}{0.48,0.40,0.93}
\definecolor{SlateBlue3}{rgb}{0.41,0.35,0.80}
\definecolor{SlateBlue4}{rgb}{0.28,0.24,0.55}
\definecolor{SlateBlue}{rgb}{0.42,0.35,0.80}
\definecolor{SlateGray1}{rgb}{0.78,0.89,1.00}
\definecolor{SlateGray2}{rgb}{0.73,0.83,0.93}
\definecolor{SlateGray3}{rgb}{0.62,0.71,0.80}
\definecolor{SlateGray4}{rgb}{0.42,0.48,0.55}
\definecolor{SlateGray}{rgb}{0.44,0.50,0.56}
\definecolor{SlateGrey}{rgb}{0.44,0.50,0.56}
\definecolor{SpringGreen1}{rgb}{0.00,1.00,0.50}
\definecolor{SpringGreen2}{rgb}{0.00,0.93,0.46}
\definecolor{SpringGreen3}{rgb}{0.00,0.80,0.40}
\definecolor{SpringGreen4}{rgb}{0.00,0.55,0.27}
\definecolor{SpringGreen}{rgb}{0.00,1.00,0.50}
\definecolor{SteelBlue1}{rgb}{0.39,0.72,1.00}
\definecolor{SteelBlue2}{rgb}{0.36,0.67,0.93}
\definecolor{SteelBlue3}{rgb}{0.31,0.58,0.80}
\definecolor{SteelBlue4}{rgb}{0.21,0.39,0.55}
\definecolor{SteelBlue}{rgb}{0.27,0.51,0.71}
\definecolor{VioletRed1}{rgb}{1.00,0.24,0.59}
\definecolor{VioletRed2}{rgb}{0.93,0.23,0.55}
\definecolor{VioletRed3}{rgb}{0.80,0.20,0.47}
\definecolor{VioletRed4}{rgb}{0.55,0.13,0.32}
\definecolor{VioletRed}{rgb}{0.82,0.13,0.56}
\definecolor{WhiteSmoke}{rgb}{0.96,0.96,0.96}
\definecolor{YellowGreen}{rgb}{0.60,0.80,0.20}
\definecolor{aliceblue}{rgb}{0.94,0.97,1.00}
\definecolor{antiquewhite}{rgb}{0.98,0.92,0.84}
\definecolor{aquamarine1}{rgb}{0.50,1.00,0.83}
\definecolor{aquamarine2}{rgb}{0.46,0.93,0.78}
\definecolor{aquamarine3}{rgb}{0.40,0.80,0.67}
\definecolor{aquamarine4}{rgb}{0.27,0.55,0.45}
\definecolor{aquamarine}{rgb}{0.50,1.00,0.83}
\definecolor{azure1}{rgb}{0.94,1.00,1.00}
\definecolor{azure2}{rgb}{0.88,0.93,0.93}
\definecolor{azure3}{rgb}{0.76,0.80,0.80}
\definecolor{azure4}{rgb}{0.51,0.55,0.55}
\definecolor{azure}{rgb}{0.94,1.00,1.00}
\definecolor{beige}{rgb}{0.96,0.96,0.86}
\definecolor{bisque1}{rgb}{1.00,0.89,0.77}
\definecolor{bisque2}{rgb}{0.93,0.84,0.72}
\definecolor{bisque3}{rgb}{0.80,0.72,0.62}
\definecolor{bisque4}{rgb}{0.55,0.49,0.42}
\definecolor{bisque}{rgb}{1.00,0.89,0.77}
\definecolor{black}{rgb}{0.00,0.00,0.00}
\definecolor{blanchedalmond}{rgb}{1.00,0.92,0.80}
\definecolor{blue1}{rgb}{0.00,0.00,1.00}
\definecolor{blue2}{rgb}{0.00,0.00,0.93}
\definecolor{blue3}{rgb}{0.00,0.00,0.80}
\definecolor{blue4}{rgb}{0.00,0.00,0.55}
\definecolor{blueviolet}{rgb}{0.54,0.17,0.89}
\definecolor{blue}{rgb}{0.00,0.00,1.00}
\definecolor{brown1}{rgb}{1.00,0.25,0.25}
\definecolor{brown2}{rgb}{0.93,0.23,0.23}
\definecolor{brown3}{rgb}{0.80,0.20,0.20}
\definecolor{brown4}{rgb}{0.55,0.14,0.14}
\definecolor{brown}{rgb}{0.65,0.16,0.16}
\definecolor{burlywood1}{rgb}{1.00,0.83,0.61}
\definecolor{burlywood2}{rgb}{0.93,0.77,0.57}
\definecolor{burlywood3}{rgb}{0.80,0.67,0.49}
\definecolor{burlywood4}{rgb}{0.55,0.45,0.33}
\definecolor{burlywood}{rgb}{0.87,0.72,0.53}
\definecolor{cadetblue}{rgb}{0.37,0.62,0.63}
\definecolor{chartreuse1}{rgb}{0.50,1.00,0.00}
\definecolor{chartreuse2}{rgb}{0.46,0.93,0.00}
\definecolor{chartreuse3}{rgb}{0.40,0.80,0.00}
\definecolor{chartreuse4}{rgb}{0.27,0.55,0.00}
\definecolor{chartreuse}{rgb}{0.50,1.00,0.00}
\definecolor{chocolate1}{rgb}{1.00,0.50,0.14}
\definecolor{chocolate2}{rgb}{0.93,0.46,0.13}
\definecolor{chocolate3}{rgb}{0.80,0.40,0.11}
\definecolor{chocolate4}{rgb}{0.55,0.27,0.07}
\definecolor{chocolate}{rgb}{0.82,0.41,0.12}
\definecolor{coral1}{rgb}{1.00,0.45,0.34}
\definecolor{coral2}{rgb}{0.93,0.42,0.31}
\definecolor{coral3}{rgb}{0.80,0.36,0.27}
\definecolor{coral4}{rgb}{0.55,0.24,0.18}
\definecolor{coral}{rgb}{1.00,0.50,0.31}
\definecolor{cornflowerblue}{rgb}{0.39,0.58,0.93}
\definecolor{cornsilk1}{rgb}{1.00,0.97,0.86}
\definecolor{cornsilk2}{rgb}{0.93,0.91,0.80}
\definecolor{cornsilk3}{rgb}{0.80,0.78,0.69}
\definecolor{cornsilk4}{rgb}{0.55,0.53,0.47}
\definecolor{cornsilk}{rgb}{1.00,0.97,0.86}
\definecolor{cyan1}{rgb}{0.00,1.00,1.00}
\definecolor{cyan2}{rgb}{0.00,0.93,0.93}
\definecolor{cyan3}{rgb}{0.00,0.80,0.80}
\definecolor{cyan4}{rgb}{0.00,0.55,0.55}
\definecolor{cyan}{rgb}{0.00,1.00,1.00}
\definecolor{darkblue}{rgb}{0.00,0.00,0.55}
\definecolor{darkcyan}{rgb}{0.00,0.55,0.55}
\definecolor{darkgoldenrod}{rgb}{0.72,0.53,0.04}
\definecolor{darkgray}{rgb}{0.66,0.66,0.66}
\definecolor{darkgreen}{rgb}{0.00,0.39,0.00}
\definecolor{darkgrey}{rgb}{0.66,0.66,0.66}
\definecolor{darkkhaki}{rgb}{0.74,0.72,0.42}
\definecolor{darkmagenta}{rgb}{0.55,0.00,0.55}
\definecolor{darkolive}{rgb}{0.33,0.42,0.18}
\definecolor{darkorange}{rgb}{1.00,0.55,0.00}
\definecolor{darkorchid}{rgb}{0.60,0.20,0.80}
\definecolor{darkred}{rgb}{0.55,0.00,0.00}
\definecolor{darksalmon}{rgb}{0.91,0.59,0.48}
\definecolor{darksea}{rgb}{0.56,0.74,0.56}
\definecolor{darkslate}{rgb}{0.18,0.31,0.31}
\definecolor{darkslate}{rgb}{0.18,0.31,0.31}
\definecolor{darkslate}{rgb}{0.28,0.24,0.55}
\definecolor{darkturquoise}{rgb}{0.00,0.81,0.82}
\definecolor{darkviolet}{rgb}{0.58,0.00,0.83}
\definecolor{deeppink}{rgb}{1.00,0.08,0.58}
\definecolor{deepsky}{rgb}{0.00,0.75,1.00}
\definecolor{dimgray}{rgb}{0.41,0.41,0.41}
\definecolor{dimgrey}{rgb}{0.41,0.41,0.41}
\definecolor{dodgerblue}{rgb}{0.12,0.56,1.00}
\definecolor{firebrick1}{rgb}{1.00,0.19,0.19}
\definecolor{firebrick2}{rgb}{0.93,0.17,0.17}
\definecolor{firebrick3}{rgb}{0.80,0.15,0.15}
\definecolor{firebrick4}{rgb}{0.55,0.10,0.10}
\definecolor{firebrick}{rgb}{0.70,0.13,0.13}
\definecolor{floralwhite}{rgb}{1.00,0.98,0.94}
\definecolor{forestgreen}{rgb}{0.13,0.55,0.13}
\definecolor{gainsboro}{rgb}{0.86,0.86,0.86}
\definecolor{ghostwhite}{rgb}{0.97,0.97,1.00}
\definecolor{gold1}{rgb}{1.00,0.84,0.00}
\definecolor{gold2}{rgb}{0.93,0.79,0.00}
\definecolor{gold3}{rgb}{0.80,0.68,0.00}
\definecolor{gold4}{rgb}{0.55,0.46,0.00}
\definecolor{goldenrod1}{rgb}{1.00,0.76,0.15}
\definecolor{goldenrod2}{rgb}{0.93,0.71,0.13}
\definecolor{goldenrod3}{rgb}{0.80,0.61,0.11}
\definecolor{goldenrod4}{rgb}{0.55,0.41,0.08}
\definecolor{goldenrod}{rgb}{0.85,0.65,0.13}
\definecolor{gold}{rgb}{1.00,0.84,0.00}
\definecolor{gray0}{rgb}{0.00,0.00,0.00}
\definecolor{gray100}{rgb}{1.00,1.00,1.00}
\definecolor{gray10}{rgb}{0.10,0.10,0.10}
\definecolor{gray11}{rgb}{0.11,0.11,0.11}
\definecolor{gray12}{rgb}{0.12,0.12,0.12}
\definecolor{gray13}{rgb}{0.13,0.13,0.13}
\definecolor{gray14}{rgb}{0.14,0.14,0.14}
\definecolor{gray15}{rgb}{0.15,0.15,0.15}
\definecolor{gray16}{rgb}{0.16,0.16,0.16}
\definecolor{gray17}{rgb}{0.17,0.17,0.17}
\definecolor{gray18}{rgb}{0.18,0.18,0.18}
\definecolor{gray19}{rgb}{0.19,0.19,0.19}
\definecolor{gray1}{rgb}{0.01,0.01,0.01}
\definecolor{gray20}{rgb}{0.20,0.20,0.20}
\definecolor{gray21}{rgb}{0.21,0.21,0.21}
\definecolor{gray22}{rgb}{0.22,0.22,0.22}
\definecolor{gray23}{rgb}{0.23,0.23,0.23}
\definecolor{gray24}{rgb}{0.24,0.24,0.24}
\definecolor{gray25}{rgb}{0.25,0.25,0.25}
\definecolor{gray26}{rgb}{0.26,0.26,0.26}
\definecolor{gray27}{rgb}{0.27,0.27,0.27}
\definecolor{gray28}{rgb}{0.28,0.28,0.28}
\definecolor{gray29}{rgb}{0.29,0.29,0.29}
\definecolor{gray2}{rgb}{0.02,0.02,0.02}
\definecolor{gray30}{rgb}{0.30,0.30,0.30}
\definecolor{gray31}{rgb}{0.31,0.31,0.31}
\definecolor{gray32}{rgb}{0.32,0.32,0.32}
\definecolor{gray33}{rgb}{0.33,0.33,0.33}
\definecolor{gray34}{rgb}{0.34,0.34,0.34}
\definecolor{gray35}{rgb}{0.35,0.35,0.35}
\definecolor{gray36}{rgb}{0.36,0.36,0.36}
\definecolor{gray37}{rgb}{0.37,0.37,0.37}
\definecolor{gray38}{rgb}{0.38,0.38,0.38}
\definecolor{gray39}{rgb}{0.39,0.39,0.39}
\definecolor{gray3}{rgb}{0.03,0.03,0.03}
\definecolor{gray40}{rgb}{0.40,0.40,0.40}
\definecolor{gray41}{rgb}{0.41,0.41,0.41}
\definecolor{gray42}{rgb}{0.42,0.42,0.42}
\definecolor{gray43}{rgb}{0.43,0.43,0.43}
\definecolor{gray44}{rgb}{0.44,0.44,0.44}
\definecolor{gray45}{rgb}{0.45,0.45,0.45}
\definecolor{gray46}{rgb}{0.46,0.46,0.46}
\definecolor{gray47}{rgb}{0.47,0.47,0.47}
\definecolor{gray48}{rgb}{0.48,0.48,0.48}
\definecolor{gray49}{rgb}{0.49,0.49,0.49}
\definecolor{gray4}{rgb}{0.04,0.04,0.04}
\definecolor{gray50}{rgb}{0.50,0.50,0.50}
\definecolor{gray51}{rgb}{0.51,0.51,0.51}
\definecolor{gray52}{rgb}{0.52,0.52,0.52}
\definecolor{gray53}{rgb}{0.53,0.53,0.53}
\definecolor{gray54}{rgb}{0.54,0.54,0.54}
\definecolor{gray55}{rgb}{0.55,0.55,0.55}
\definecolor{gray56}{rgb}{0.56,0.56,0.56}
\definecolor{gray57}{rgb}{0.57,0.57,0.57}
\definecolor{gray58}{rgb}{0.58,0.58,0.58}
\definecolor{gray59}{rgb}{0.59,0.59,0.59}
\definecolor{gray5}{rgb}{0.05,0.05,0.05}
\definecolor{gray60}{rgb}{0.60,0.60,0.60}
\definecolor{gray61}{rgb}{0.61,0.61,0.61}
\definecolor{gray62}{rgb}{0.62,0.62,0.62}
\definecolor{gray63}{rgb}{0.63,0.63,0.63}
\definecolor{gray64}{rgb}{0.64,0.64,0.64}
\definecolor{gray65}{rgb}{0.65,0.65,0.65}
\definecolor{gray66}{rgb}{0.66,0.66,0.66}
\definecolor{gray67}{rgb}{0.67,0.67,0.67}
\definecolor{gray68}{rgb}{0.68,0.68,0.68}
\definecolor{gray69}{rgb}{0.69,0.69,0.69}
\definecolor{gray6}{rgb}{0.06,0.06,0.06}
\definecolor{gray70}{rgb}{0.70,0.70,0.70}
\definecolor{gray71}{rgb}{0.71,0.71,0.71}
\definecolor{gray72}{rgb}{0.72,0.72,0.72}
\definecolor{gray73}{rgb}{0.73,0.73,0.73}
\definecolor{gray74}{rgb}{0.74,0.74,0.74}
\definecolor{gray75}{rgb}{0.75,0.75,0.75}
\definecolor{gray76}{rgb}{0.76,0.76,0.76}
\definecolor{gray77}{rgb}{0.77,0.77,0.77}
\definecolor{gray78}{rgb}{0.78,0.78,0.78}
\definecolor{gray79}{rgb}{0.79,0.79,0.79}
\definecolor{gray7}{rgb}{0.07,0.07,0.07}
\definecolor{gray80}{rgb}{0.80,0.80,0.80}
\definecolor{gray81}{rgb}{0.81,0.81,0.81}
\definecolor{gray82}{rgb}{0.82,0.82,0.82}
\definecolor{gray83}{rgb}{0.83,0.83,0.83}
\definecolor{gray84}{rgb}{0.84,0.84,0.84}
\definecolor{gray85}{rgb}{0.85,0.85,0.85}
\definecolor{gray86}{rgb}{0.86,0.86,0.86}
\definecolor{gray87}{rgb}{0.87,0.87,0.87}
\definecolor{gray88}{rgb}{0.88,0.88,0.88}
\definecolor{gray89}{rgb}{0.89,0.89,0.89}
\definecolor{gray8}{rgb}{0.08,0.08,0.08}
\definecolor{gray90}{rgb}{0.90,0.90,0.90}
\definecolor{gray91}{rgb}{0.91,0.91,0.91}
\definecolor{gray92}{rgb}{0.92,0.92,0.92}
\definecolor{gray93}{rgb}{0.93,0.93,0.93}
\definecolor{gray94}{rgb}{0.94,0.94,0.94}
\definecolor{gray95}{rgb}{0.95,0.95,0.95}
\definecolor{gray96}{rgb}{0.96,0.96,0.96}
\definecolor{gray97}{rgb}{0.97,0.97,0.97}
\definecolor{gray98}{rgb}{0.98,0.98,0.98}
\definecolor{gray99}{rgb}{0.99,0.99,0.99}
\definecolor{gray9}{rgb}{0.09,0.09,0.09}
\definecolor{gray}{rgb}{0.75,0.75,0.75}
\definecolor{green1}{rgb}{0.00,1.00,0.00}
\definecolor{green2}{rgb}{0.00,0.93,0.00}
\definecolor{green3}{rgb}{0.00,0.80,0.00}
\definecolor{green4}{rgb}{0.00,0.55,0.00}
\definecolor{greenyellow}{rgb}{0.68,1.00,0.18}
\definecolor{green}{rgb}{0.00,1.00,0.00}
\definecolor{grey0}{rgb}{0.00,0.00,0.00}
\definecolor{grey100}{rgb}{1.00,1.00,1.00}
\definecolor{grey10}{rgb}{0.10,0.10,0.10}
\definecolor{grey11}{rgb}{0.11,0.11,0.11}
\definecolor{grey12}{rgb}{0.12,0.12,0.12}
\definecolor{grey13}{rgb}{0.13,0.13,0.13}
\definecolor{grey14}{rgb}{0.14,0.14,0.14}
\definecolor{grey15}{rgb}{0.15,0.15,0.15}
\definecolor{grey16}{rgb}{0.16,0.16,0.16}
\definecolor{grey17}{rgb}{0.17,0.17,0.17}
\definecolor{grey18}{rgb}{0.18,0.18,0.18}
\definecolor{grey19}{rgb}{0.19,0.19,0.19}
\definecolor{grey1}{rgb}{0.01,0.01,0.01}
\definecolor{grey20}{rgb}{0.20,0.20,0.20}
\definecolor{grey21}{rgb}{0.21,0.21,0.21}
\definecolor{grey22}{rgb}{0.22,0.22,0.22}
\definecolor{grey23}{rgb}{0.23,0.23,0.23}
\definecolor{grey24}{rgb}{0.24,0.24,0.24}
\definecolor{grey25}{rgb}{0.25,0.25,0.25}
\definecolor{grey26}{rgb}{0.26,0.26,0.26}
\definecolor{grey27}{rgb}{0.27,0.27,0.27}
\definecolor{grey28}{rgb}{0.28,0.28,0.28}
\definecolor{grey29}{rgb}{0.29,0.29,0.29}
\definecolor{grey2}{rgb}{0.02,0.02,0.02}
\definecolor{grey30}{rgb}{0.30,0.30,0.30}
\definecolor{grey31}{rgb}{0.31,0.31,0.31}
\definecolor{grey32}{rgb}{0.32,0.32,0.32}
\definecolor{grey33}{rgb}{0.33,0.33,0.33}
\definecolor{grey34}{rgb}{0.34,0.34,0.34}
\definecolor{grey35}{rgb}{0.35,0.35,0.35}
\definecolor{grey36}{rgb}{0.36,0.36,0.36}
\definecolor{grey37}{rgb}{0.37,0.37,0.37}
\definecolor{grey38}{rgb}{0.38,0.38,0.38}
\definecolor{grey39}{rgb}{0.39,0.39,0.39}
\definecolor{grey3}{rgb}{0.03,0.03,0.03}
\definecolor{grey40}{rgb}{0.40,0.40,0.40}
\definecolor{grey41}{rgb}{0.41,0.41,0.41}
\definecolor{grey42}{rgb}{0.42,0.42,0.42}
\definecolor{grey43}{rgb}{0.43,0.43,0.43}
\definecolor{grey44}{rgb}{0.44,0.44,0.44}
\definecolor{grey45}{rgb}{0.45,0.45,0.45}
\definecolor{grey46}{rgb}{0.46,0.46,0.46}
\definecolor{grey47}{rgb}{0.47,0.47,0.47}
\definecolor{grey48}{rgb}{0.48,0.48,0.48}
\definecolor{grey49}{rgb}{0.49,0.49,0.49}
\definecolor{grey4}{rgb}{0.04,0.04,0.04}
\definecolor{grey50}{rgb}{0.50,0.50,0.50}
\definecolor{grey51}{rgb}{0.51,0.51,0.51}
\definecolor{grey52}{rgb}{0.52,0.52,0.52}
\definecolor{grey53}{rgb}{0.53,0.53,0.53}
\definecolor{grey54}{rgb}{0.54,0.54,0.54}
\definecolor{grey55}{rgb}{0.55,0.55,0.55}
\definecolor{grey56}{rgb}{0.56,0.56,0.56}
\definecolor{grey57}{rgb}{0.57,0.57,0.57}
\definecolor{grey58}{rgb}{0.58,0.58,0.58}
\definecolor{grey59}{rgb}{0.59,0.59,0.59}
\definecolor{grey5}{rgb}{0.05,0.05,0.05}
\definecolor{grey60}{rgb}{0.60,0.60,0.60}
\definecolor{grey61}{rgb}{0.61,0.61,0.61}
\definecolor{grey62}{rgb}{0.62,0.62,0.62}
\definecolor{grey63}{rgb}{0.63,0.63,0.63}
\definecolor{grey64}{rgb}{0.64,0.64,0.64}
\definecolor{grey65}{rgb}{0.65,0.65,0.65}
\definecolor{grey66}{rgb}{0.66,0.66,0.66}
\definecolor{grey67}{rgb}{0.67,0.67,0.67}
\definecolor{grey68}{rgb}{0.68,0.68,0.68}
\definecolor{grey69}{rgb}{0.69,0.69,0.69}
\definecolor{grey6}{rgb}{0.06,0.06,0.06}
\definecolor{grey70}{rgb}{0.70,0.70,0.70}
\definecolor{grey71}{rgb}{0.71,0.71,0.71}
\definecolor{grey72}{rgb}{0.72,0.72,0.72}
\definecolor{grey73}{rgb}{0.73,0.73,0.73}
\definecolor{grey74}{rgb}{0.74,0.74,0.74}
\definecolor{grey75}{rgb}{0.75,0.75,0.75}
\definecolor{grey76}{rgb}{0.76,0.76,0.76}
\definecolor{grey77}{rgb}{0.77,0.77,0.77}
\definecolor{grey78}{rgb}{0.78,0.78,0.78}
\definecolor{grey79}{rgb}{0.79,0.79,0.79}
\definecolor{grey7}{rgb}{0.07,0.07,0.07}
\definecolor{grey80}{rgb}{0.80,0.80,0.80}
\definecolor{grey81}{rgb}{0.81,0.81,0.81}
\definecolor{grey82}{rgb}{0.82,0.82,0.82}
\definecolor{grey83}{rgb}{0.83,0.83,0.83}
\definecolor{grey84}{rgb}{0.84,0.84,0.84}
\definecolor{grey85}{rgb}{0.85,0.85,0.85}
\definecolor{grey86}{rgb}{0.86,0.86,0.86}
\definecolor{grey87}{rgb}{0.87,0.87,0.87}
\definecolor{grey88}{rgb}{0.88,0.88,0.88}
\definecolor{grey89}{rgb}{0.89,0.89,0.89}
\definecolor{grey8}{rgb}{0.08,0.08,0.08}
\definecolor{grey90}{rgb}{0.90,0.90,0.90}
\definecolor{grey91}{rgb}{0.91,0.91,0.91}
\definecolor{grey92}{rgb}{0.92,0.92,0.92}
\definecolor{grey93}{rgb}{0.93,0.93,0.93}
\definecolor{grey94}{rgb}{0.94,0.94,0.94}
\definecolor{grey95}{rgb}{0.95,0.95,0.95}
\definecolor{grey96}{rgb}{0.96,0.96,0.96}
\definecolor{grey97}{rgb}{0.97,0.97,0.97}
\definecolor{grey98}{rgb}{0.98,0.98,0.98}
\definecolor{grey99}{rgb}{0.99,0.99,0.99}
\definecolor{grey9}{rgb}{0.09,0.09,0.09}
\definecolor{grey}{rgb}{0.75,0.75,0.75}
\definecolor{honeydew1}{rgb}{0.94,1.00,0.94}
\definecolor{honeydew2}{rgb}{0.88,0.93,0.88}
\definecolor{honeydew3}{rgb}{0.76,0.80,0.76}
\definecolor{honeydew4}{rgb}{0.51,0.55,0.51}
\definecolor{honeydew}{rgb}{0.94,1.00,0.94}
\definecolor{hotpink}{rgb}{1.00,0.41,0.71}
\definecolor{indianred}{rgb}{0.80,0.36,0.36}
\definecolor{ivory1}{rgb}{1.00,1.00,0.94}
\definecolor{ivory2}{rgb}{0.93,0.93,0.88}
\definecolor{ivory3}{rgb}{0.80,0.80,0.76}
\definecolor{ivory4}{rgb}{0.55,0.55,0.51}
\definecolor{ivory}{rgb}{1.00,1.00,0.94}
\definecolor{khaki1}{rgb}{1.00,0.96,0.56}
\definecolor{khaki2}{rgb}{0.93,0.90,0.52}
\definecolor{khaki3}{rgb}{0.80,0.78,0.45}
\definecolor{khaki4}{rgb}{0.55,0.53,0.31}
\definecolor{khaki}{rgb}{0.94,0.90,0.55}
\definecolor{lavenderblush}{rgb}{1.00,0.94,0.96}
\definecolor{lavender}{rgb}{0.90,0.90,0.98}
\definecolor{lawngreen}{rgb}{0.49,0.99,0.00}
\definecolor{lemonchiffon}{rgb}{1.00,0.98,0.80}
\definecolor{lightblue}{rgb}{0.68,0.85,0.90}
\definecolor{lightcoral}{rgb}{0.94,0.50,0.50}
\definecolor{lightcyan}{rgb}{0.88,1.00,1.00}
\definecolor{lightgoldenrod}{rgb}{0.93,0.87,0.51}
\definecolor{lightgoldenrod}{rgb}{0.98,0.98,0.82}
\definecolor{lightgray}{rgb}{0.83,0.83,0.83}
\definecolor{lightgreen}{rgb}{0.56,0.93,0.56}
\definecolor{lightgrey}{rgb}{0.83,0.83,0.83}
\definecolor{lightpink}{rgb}{1.00,0.71,0.76}
\definecolor{lightsalmon}{rgb}{1.00,0.63,0.48}
\definecolor{lightsea}{rgb}{0.13,0.70,0.67}
\definecolor{lightsky}{rgb}{0.53,0.81,0.98}
\definecolor{lightslate}{rgb}{0.47,0.53,0.60}
\definecolor{lightslate}{rgb}{0.47,0.53,0.60}
\definecolor{lightslate}{rgb}{0.52,0.44,1.00}
\definecolor{lightsteel}{rgb}{0.69,0.77,0.87}
\definecolor{lightyellow}{rgb}{1.00,1.00,0.88}
\definecolor{limegreen}{rgb}{0.20,0.80,0.20}
\definecolor{linen}{rgb}{0.98,0.94,0.90}
\definecolor{magenta1}{rgb}{1.00,0.00,1.00}
\definecolor{magenta2}{rgb}{0.93,0.00,0.93}
\definecolor{magenta3}{rgb}{0.80,0.00,0.80}
\definecolor{magenta4}{rgb}{0.55,0.00,0.55}
\definecolor{magenta}{rgb}{1.00,0.00,1.00}
\definecolor{maroon1}{rgb}{1.00,0.20,0.70}
\definecolor{maroon2}{rgb}{0.93,0.19,0.65}
\definecolor{maroon3}{rgb}{0.80,0.16,0.56}
\definecolor{maroon4}{rgb}{0.55,0.11,0.38}
\definecolor{maroon}{rgb}{0.69,0.19,0.38}
\definecolor{mediumaquamarine}{rgb}{0.40,0.80,0.67}
\definecolor{mediumblue}{rgb}{0.00,0.00,0.80}
\definecolor{mediumorchid}{rgb}{0.73,0.33,0.83}
\definecolor{mediumpurple}{rgb}{0.58,0.44,0.86}
\definecolor{mediumsea}{rgb}{0.24,0.70,0.44}
\definecolor{mediumslate}{rgb}{0.48,0.41,0.93}
\definecolor{mediumspring}{rgb}{0.00,0.98,0.60}
\definecolor{mediumturquoise}{rgb}{0.28,0.82,0.80}
\definecolor{mediumviolet}{rgb}{0.78,0.08,0.52}
\definecolor{midnightblue}{rgb}{0.10,0.10,0.44}
\definecolor{mintcream}{rgb}{0.96,1.00,0.98}
\definecolor{mistyrose}{rgb}{1.00,0.89,0.88}
\definecolor{moccasin}{rgb}{1.00,0.89,0.71}
\definecolor{navajowhite}{rgb}{1.00,0.87,0.68}
\definecolor{navyblue}{rgb}{0.00,0.00,0.50}
\definecolor{navy}{rgb}{0.00,0.00,0.50}
\definecolor{oldlace}{rgb}{0.99,0.96,0.90}
\definecolor{olivedrab}{rgb}{0.42,0.56,0.14}
\definecolor{orange1}{rgb}{1.00,0.65,0.00}
\definecolor{orange2}{rgb}{0.93,0.60,0.00}
\definecolor{orange3}{rgb}{0.80,0.52,0.00}
\definecolor{orange4}{rgb}{0.55,0.35,0.00}
\definecolor{orangered}{rgb}{1.00,0.27,0.00}
\definecolor{orange}{rgb}{1.00,0.65,0.00}
\definecolor{orchid1}{rgb}{1.00,0.51,0.98}
\definecolor{orchid2}{rgb}{0.93,0.48,0.91}
\definecolor{orchid3}{rgb}{0.80,0.41,0.79}
\definecolor{orchid4}{rgb}{0.55,0.28,0.54}
\definecolor{orchid}{rgb}{0.85,0.44,0.84}
\definecolor{palegoldenrod}{rgb}{0.93,0.91,0.67}
\definecolor{palegreen}{rgb}{0.60,0.98,0.60}
\definecolor{paleturquoise}{rgb}{0.69,0.93,0.93}
\definecolor{paleviolet}{rgb}{0.86,0.44,0.58}
\definecolor{papayawhip}{rgb}{1.00,0.94,0.84}
\definecolor{peachpuff}{rgb}{1.00,0.85,0.73}
\definecolor{peru}{rgb}{0.80,0.52,0.25}
\definecolor{pink1}{rgb}{1.00,0.71,0.77}
\definecolor{pink2}{rgb}{0.93,0.66,0.72}
\definecolor{pink3}{rgb}{0.80,0.57,0.62}
\definecolor{pink4}{rgb}{0.55,0.39,0.42}
\definecolor{pink}{rgb}{1.00,0.75,0.80}
\definecolor{plum1}{rgb}{1.00,0.73,1.00}
\definecolor{plum2}{rgb}{0.93,0.68,0.93}
\definecolor{plum3}{rgb}{0.80,0.59,0.80}
\definecolor{plum4}{rgb}{0.55,0.40,0.55}
\definecolor{plum}{rgb}{0.87,0.63,0.87}
\definecolor{powderblue}{rgb}{0.69,0.88,0.90}
\definecolor{purple1}{rgb}{0.61,0.19,1.00}
\definecolor{purple2}{rgb}{0.57,0.17,0.93}
\definecolor{purple3}{rgb}{0.49,0.15,0.80}
\definecolor{purple4}{rgb}{0.33,0.10,0.55}
\definecolor{purple}{rgb}{0.63,0.13,0.94}
\definecolor{red1}{rgb}{1.00,0.00,0.00}
\definecolor{red2}{rgb}{0.93,0.00,0.00}
\definecolor{red3}{rgb}{0.80,0.00,0.00}
\definecolor{red4}{rgb}{0.55,0.00,0.00}
\definecolor{red}{rgb}{1.00,0.00,0.00}
\definecolor{rosybrown}{rgb}{0.74,0.56,0.56}
\definecolor{royalblue}{rgb}{0.25,0.41,0.88}
\definecolor{saddlebrown}{rgb}{0.55,0.27,0.07}
\definecolor{salmon1}{rgb}{1.00,0.55,0.41}
\definecolor{salmon2}{rgb}{0.93,0.51,0.38}
\definecolor{salmon3}{rgb}{0.80,0.44,0.33}
\definecolor{salmon4}{rgb}{0.55,0.30,0.22}
\definecolor{salmon}{rgb}{0.98,0.50,0.45}
\definecolor{sandybrown}{rgb}{0.96,0.64,0.38}
\definecolor{seagreen}{rgb}{0.18,0.55,0.34}
\definecolor{seashell1}{rgb}{1.00,0.96,0.93}
\definecolor{seashell2}{rgb}{0.93,0.90,0.87}
\definecolor{seashell3}{rgb}{0.80,0.77,0.75}
\definecolor{seashell4}{rgb}{0.55,0.53,0.51}
\definecolor{seashell}{rgb}{1.00,0.96,0.93}
\definecolor{sienna1}{rgb}{1.00,0.51,0.28}
\definecolor{sienna2}{rgb}{0.93,0.47,0.26}
\definecolor{sienna3}{rgb}{0.80,0.41,0.22}
\definecolor{sienna4}{rgb}{0.55,0.28,0.15}
\definecolor{sienna}{rgb}{0.63,0.32,0.18}
\definecolor{skyblue}{rgb}{0.53,0.81,0.92}
\definecolor{slateblue}{rgb}{0.42,0.35,0.80}
\definecolor{slategray}{rgb}{0.44,0.50,0.56}
\definecolor{slategrey}{rgb}{0.44,0.50,0.56}
\definecolor{snow1}{rgb}{1.00,0.98,0.98}
\definecolor{snow2}{rgb}{0.93,0.91,0.91}
\definecolor{snow3}{rgb}{0.80,0.79,0.79}
\definecolor{snow4}{rgb}{0.55,0.54,0.54}
\definecolor{snow}{rgb}{1.00,0.98,0.98}
\definecolor{springgreen}{rgb}{0.00,1.00,0.50}
\definecolor{steelblue}{rgb}{0.27,0.51,0.71}
\definecolor{tan1}{rgb}{1.00,0.65,0.31}
\definecolor{tan2}{rgb}{0.93,0.60,0.29}
\definecolor{tan3}{rgb}{0.80,0.52,0.25}
\definecolor{tan4}{rgb}{0.55,0.35,0.17}
\definecolor{tan}{rgb}{0.82,0.71,0.55}
\definecolor{thistle1}{rgb}{1.00,0.88,1.00}
\definecolor{thistle2}{rgb}{0.93,0.82,0.93}
\definecolor{thistle3}{rgb}{0.80,0.71,0.80}
\definecolor{thistle4}{rgb}{0.55,0.48,0.55}
\definecolor{thistle}{rgb}{0.85,0.75,0.85}
\definecolor{tomato1}{rgb}{1.00,0.39,0.28}
\definecolor{tomato2}{rgb}{0.93,0.36,0.26}
\definecolor{tomato3}{rgb}{0.80,0.31,0.22}
\definecolor{tomato4}{rgb}{0.55,0.21,0.15}
\definecolor{tomato}{rgb}{1.00,0.39,0.28}
\definecolor{turquoise1}{rgb}{0.00,0.96,1.00}
\definecolor{turquoise2}{rgb}{0.00,0.90,0.93}
\definecolor{turquoise3}{rgb}{0.00,0.77,0.80}
\definecolor{turquoise4}{rgb}{0.00,0.53,0.55}
\definecolor{turquoise}{rgb}{0.25,0.88,0.82}
\definecolor{violetred}{rgb}{0.82,0.13,0.56}
\definecolor{violet}{rgb}{0.93,0.51,0.93}
\definecolor{wheat1}{rgb}{1.00,0.91,0.73}
\definecolor{wheat2}{rgb}{0.93,0.85,0.68}
\definecolor{wheat3}{rgb}{0.80,0.73,0.59}
\definecolor{wheat4}{rgb}{0.55,0.49,0.40}
\definecolor{wheat}{rgb}{0.96,0.87,0.70}
\definecolor{whitesmoke}{rgb}{0.96,0.96,0.96}
\definecolor{white}{rgb}{1.00,1.00,1.00}
\definecolor{yellow1}{rgb}{1.00,1.00,0.00}
\definecolor{yellow2}{rgb}{0.93,0.93,0.00}
\definecolor{yellow3}{rgb}{0.80,0.80,0.00}
\definecolor{yellow4}{rgb}{0.55,0.55,0.00}
\definecolor{yellowgreen}{rgb}{0.60,0.80,0.20}
\definecolor{yellow}{rgb}{1.00,1.00,0.00}

\usepackage{amsmath}
\usepackage{amssymb}
\usepackage[dvips]{graphicx}
\usepackage{amsmath,amsfonts,amssymb,aas_macros}

\usepackage[normalem]{ulem} 
\usepackage{epstopdf}
\usepackage{epsfig}

\usepackage{hyperref}
\usepackage{tikz}
\usepackage{listings}
\usepackage[normalem]{ulem} 

\usepackage{algpseudocode}
\usepackage{algorithm}
\usepackage{lscape}
\usepackage{longtable,booktabs}
\usepackage{caption}

\newcommand{\gadget}{\textsc{gadget}}
\newcommand{\pthalos}{\textsc{PThalos}}
\newcommand{\patchy}{\textsc{patchy}}
\newcommand{\cola}{\textsc{cola}}
\newcommand{\qpm}{\textsc{qpm}}
\newcommand{\nbody}{$N$-body}
\newcommand{\ezmocks}{\textsc{EZmocks}}
\newcommand{\halogen}{\textsc{halogen}}
\newcommand{\pinocchio}{\textsc{pinocchio}}
\newcommand{\hmfcalc}{\textsc{HMFcalc}}
\newcommand{\ahf}{\textsc{ahf}}
\newcommand{\fof}{\textsc{fof}}
\newcommand{\goliat}{\textsc{goliat}}
\newcommand{\bigmd}{\textsc{BigMultiDark}}
\newcommand{\Eq}[1]{Eq.~(\ref{#1})}
\newcommand{\Fig}[1]{Figure~\ref{#1}}

\newcommand{\Sec}[1]{\S\ref{#1}}
\newcommand{\Tab}[1]{Table~\ref{#1}}

\newcommand{\hMsun}{\mbox{$h^{-1}M_\odot$}}
\newcommand{\hMpc}{{\ifmmode{h^{-1}{\rm Mpc}}\else{$h^{-1}$Mpc}\fi}}
\newcommand{\hkpc}{{\ifmmode{h^{-1}{\rm kpc}}\else{$h^{-1}$kpc}\fi}}
\newcommand{\rc}{\rho_{\rm cell}}

\title[\halogen]{\vskip -0.62in
\begin{minipage}{7.03 in}
\begin{flushright}
{\rm \small IFT-UAM/CSIC-14-128}
\end{flushright}
\end{minipage}\\
\vskip 1.0in
HALOGEN: A tool for fast generation of mock halo catalogues
}

\author[Avila et al.]{Santiago~Avila$^{1,2}$\thanks{santiago.avila@uam.es}, Steven~G.~Murray$^{3,4}$\thanks{steven.murray@icrar.org}, Alexander~Knebe$^{1},$ Chris~Power$^{3,4}$, 
\newauthor
         Aaron~S.~G.~Robotham$^{3}$, Juan~Garcia-Bellido$^{1,2}$\\
	  $^{1}$Departamento de F\'isica Te\'orica, M\'odulo C-15, Facultad de Ciencias, Universidad Aut\'onoma de Madrid, 28049 Cantoblanco, Madrid, Spain\\
	$^{2}$Instituto de F\'isica Te\'orica, UAM-CSIC, Universidad Autonoma de Madrid, 28049 Cantoblanco, Madrid, Spain\\
        $^3$ICRAR, University of Western Australia, 35 Stirling Highway,
        Crawley, Western Australia 6009, Australia\\
        $^4$ARC Centre of Excellence for All-Sky Astrophysics (CAASTRO)}
\date{\today}

\begin{document}

\pagerange{\pageref{firstpage}--\pageref{lastpage}} \pubyear{2013}\volume{0000}

\maketitle

\label{firstpage}

\begin{abstract}
We present a simple method of generating approximate synthetic halo catalogues: \halogen.
This method uses a combination of $2^{nd}$-order Lagrangian Perturbation Theory (2LPT) in order to generate the large-scale matter distribution, 
analytical mass functions to generate halo masses, and a single-parameter stochastic model for halo bias to position haloes. 
\halogen\ represents a simplification of  similar recently published methods.

Our method is constrained to recover the 2-point function at intermediate ($10\hMpc<r<50\hMpc$) scales, which we show is successful to within $2$ per cent.
Larger scales ($\sim100\hMpc$) are reproduced to within $15$ per cent. We compare several other statistics (e.g. power spectrum, point distribution function, redshift space distortions) 
with results from \nbody\ simulations to determine the validity 
of our method for different purposes. One of the benefits of \halogen\ is its flexibility, and we demonstrate this by showing how it can be adapted to varying 
cosmologies and simulation specifications.

A driving motivation for the development of such approximate schemes is the need to compute covariance matrices and study the systematic errors for large galaxy surveys, 
which requires thousands of simulated realisations. We discuss the applicability of our method in this context, and conclude that it is well suited to mass 
production of appropriate halo catalogues.

The code is publicly available at \url{https://github.com/savila/halogen}
\end{abstract}




\section{Introduction}
\label{sec:intro}
We have entered an observational era where it is customary for redshift surveys to map millions of galaxies in the sky with the volumes of these surveys exceeding Gpc$^3$ 
scales. 
Recent and upcoming galaxy survey projects include PAU \citep{PAU}, BOSS \citep{BOSS}, DES \citep{DES}, DESi \citep{DESi}, Euclid \citep{Euclid}, etc. 
The interpretation of such surveys demands a new generation  
of theory tools in order to better understand and interpret the large amounts of data. One important component is the need for accurate simulations of the expected results, 
to which the observations should be compared. 
However, models of large-scale-structure and the clustering of (dark matter) haloes forming in it are inherently non-linear, and require the production of simulations
based on \nbody\ calculations. Such simulations are extremely costly, and consequently very few realisations can be run for a given application. 
However,  investigating the effects of systematic errors, cosmic variance, and their interplay require many hundreds of realisations of a single simulation (e.g. BOSS survey used 600 \citep{PTHalos_MM}). These are necessary to compute covariance matrices which characterise the resultant uncertainty on the final parameters.

To mitigate this situation, many have now turned to approximate schemes in order to calculate the required realisations of the simulations. 
Early such work used the so-called log-normal realisations \citep{lognormal}, which placed particles randomly according to a log-normal distribution, given the true power spectrum. 
While this is indeed efficient, and reproduces 2-point statistics faithfully, its lack of physical motivation for the particle placement results in poor higher-order statistics, such as the 3-point function or Counts-in-Cells moments.  
Improved methods developed in the past decade include \pthalos\ \citep{PTHalos_MM,PTHalos}, \pinocchio\ \citep{pinocchio_1,pinocchio_2}, \patchy\ \citep{patchy}, \cola\ \citep{cola}, \qpm\ \citep{qpm}, 
\ezmocks\ \citep{EZmocks}, etc. For a comparison of these methods (incl. \halogen\ presented here) we refer the reader to \citet{nifty}.  

One may segregate these methods into two classes -- predictive-type methods which are required to `find' haloes in a given density field (eg. \pinocchio, \cola\  and \pthalos), and statistical-type methods which merely stochastically sample a density field to locate haloes (eg. \patchy, \qpm\ and \ezmocks). The former have the advantage of being predictive, and often not requiring an \nbody\ reference simulation for calibration, while the latter have the advantage of computational speed and resources, as the number of particles used is reduced.

We present a new (statistical-type) approximate scheme, called \halogen\, whose prime objective is to generate halo catalogues with the correct 2-point clustering and mass-dependent bias using a simple and rapid approach.

We note that statistical-type methods tend to follow a standard pattern of four steps:
\begin{enumerate}
	\item Produce a density field.
	\item Sample halo masses.
	\item Sample particles as halos with some bias.
	\item Assign halo velocities.
\end{enumerate}

In this paper we seek to abstract this pattern, providing a framework in which each step is highly modular. 
Whilst modular, \halogen\ implements default behaviour with very simple (and rapid) components -- using the popular  $2^{\rm{nd}}$-order Lagrangian Perturbation Theory (2LPT) as the gravity solver, theoretical mass functions, a single-parameter bias prescription (as opposed to 2 or more parameters for other statistical-type methods) and a direct linear transformation of the velocities. As such, \halogen\ can be rapidly calibrated, and easily extended.
In addition, we introduce physically motivated constraints for \textit{halo exclusion} and \textit{mass conservation}, which tie the individual steps together. 

In this paper we will compare the results from \halogen\ to a pair of reference \nbody\ simulations to be presented in \Sec{sec:data}. 
We introduce the general ideas of the method in \Sec{sec:method}, leaving a more detailed explanation of the spatial placement of haloes -- which we consider the essence of \halogen\ -- for \Sec{sec:halogen}. \Sec{sec:parameters} demonstrates the effects of each parameter of \halogen\, and how to optimise them. We conclude with some applications and results in \Sec{sec:applications}.

\section{The Reference Simulations}
\label{sec:data}
To tune \halogen\ to a specific cosmology, we require an \nbody\ simulation. In order to show the adaptability of \halogen\ to varying setups, 
we have not limited ourselves to a single simulation but used two with differing box size, mass resolution, and cosmology. 
Further, the reference halo catalogues have been obtained by applying two different halo finding techniques, and have different number density.
We summarise the characteristics of both reference catalogues in \Tab{tab:sims} and describe them below. 

\paragraph*{Goliat Simulation}
This simulation was run with the \gadget2 code \citep{gadget} from initial conditions generated by \textsc{2LPTic}\footnote{\url{http://cosmo.nyu.edu/roman/2LPT}} at $z=32$. It uses 
$N=512^3$ dark matter particles in a box with side length $L_{\rm box}=1000$\hMpc. The cosmological parameters used in this simulation are $\Omega_M =0.27$, 
$\Omega_\Lambda =0.73$,
$\Omega_b=0.044$, $h=0.7$, $\sigma_8=0.8$, $n_s =0.96$ yielding a mass resolution of $m_p=5.58\times 10^{11}$\hMsun.
The halo catalogue was obtained from a $z=0$ snapshot and has been generated with the halo finder \ahf\ \citep{ahf}, a spherical-overdensity (SO) algorithm. Though \ahf\ identifies subhaloes, they have been
discarded for the present analysis as these scales are too small for 2LPT to resolve.
There is a possibility of phenomenologically adding substructure in a later step using a Halo Occupation Distribution 
(HOD) prescription \citep{Skibba2009},
but we leave that to a future study.
In this catalogue we use a halo reference density of $n=2.0\cdot10^{-4} ({\rm Mpc}/h)^{-3}$

\halogen\ requires an input density field obtained from 2LPT (see \Sec{sec:2lpt}). For this purpose, we run a \textsc{2LPTic} snapshot at $z=0$ with the same 
inital condition phases as those used in \goliat.

\paragraph*{Big MultiDark Simulation\footnote{\url{http://www.cosmosim.org}}}
\bigmd\, described in \cite{bigmd}, employs the cosmology from the Planck CMB mission \citep{planck}, which for some parameters represents a significant
change with respect to the \goliat\ simulation:
$\Omega_M =0.31$, $\Omega_\Lambda =0.69$, $\Omega_b=0.048$, $h=0.68$, $\sigma_8=0.82$, $n_s =0.96$.
The halo catalogue is extracted with a Friends-of-Friends (\fof) 
\citep{Davis85} 
algorithm (which intrinsically neglects substructure) at $z=0.56$, and we choose a reference halo number density of 
$n=3.5\cdot10^{-4} (Mpc/h)^{-3}$.

Compared to \goliat, is both larger ($L_{\rm box}=2500$\hMpc) and more resolved ($N=3840^3$ particles of mass 
$m_p=2.3\times 10^{10}$\hMsun). 
It was run with L-\gadget2 from initial conditions based on the Zel'dovich Approximation (ZA) at $z=100$.
Given the large scales that it explores while resolving large numbers of haloes, it is well suited to probing the Baryon Acoustic Oscillation (BAO) peak.

For the input of \halogen\ we run \textsc{2LPTic} to $z=0.56$ with the same initial condition phases as \bigmd. The cosmology and $L_{box}$ used are the same,
but with a lower resolution of $N=1280^3$.

  \begin{table*}
        \centering
      \begin{tabular}[width=\linewidth]{cccccccccccccc}
      \hline
       Name 		& $L_{\rm box}$ & $N_{\rm part}$	& $z$ & $\Omega_b$ & $\Omega_M$ & $\Omega_\Lambda$ & $h$ & $\sigma_8$ & $n_S$ & Finder & $n$ & IC &$z_{\rm IC}$  \\
      \hline
	\goliat\ 	& $1000$	& $512^3$, $512^3$ 	& 0	& 0.044 & 0.27 & 0.69 & 0.7  & 0.8  & 0.96 & AHF & $2.0\cdot 10^{-4}$ & 2LPT & 32 \\
	\bigmd\ 	& $2500$	& $3840^3$, $1280^3$ 	& 0.56	& 0.048 & 0.31 & 0.73 & 0.68 & 0.82 & 0.96 & FOF & $3.5\cdot 10^{-4}$ & ZA & 100\\
      \hline
      \end{tabular}
      \caption{Properties of the two reference \nbody\ halo catalogues. From left to right: Side-length of the simulated cubic volume (in $\hMpc$), number of particles (for \nbody\ and \halogen),
      redshift of the snapshot, cosmological parameters (density of barions, total matter and dark energy, Hubble parameter, power spectrum normalisation and spectral index), halo finding 
      technique, halo number density (in $({\rm Mpc}/h)^{-3}$), method used to generate the initial conditions and redshift at which they were generated.} 
   \label{tab:sims}
 \end{table*}

\section{Method Outline}
\label{sec:method}
In this section we briefly outline our method, leaving a more detailed presentation of the actual modus operandi of \halogen\ for \Sec{sec:halogen}. The general algorithm consists of four (major) steps:
\begin{itemize}
	\item generate a dark matter density field,
	\item draw halo masses by sampling a halo mass function,
	\item populate the volume with haloes in the box, and
	\item assign velocities to the haloes.
\end{itemize}

We aim to de-couple each of these steps from the others as far as possible so that different algorithms may be used at each point.
The first two steps are relatively trivial, as they use pre-developed prescriptions from the literature, and we discuss these, and basic outlines of the last two steps, in this section.  

\subsection{Density Field} \label{sec:2lpt}

 \begin{figure}
 \begin{center}
   \centering
   \includegraphics[width=0.8\columnwidth]{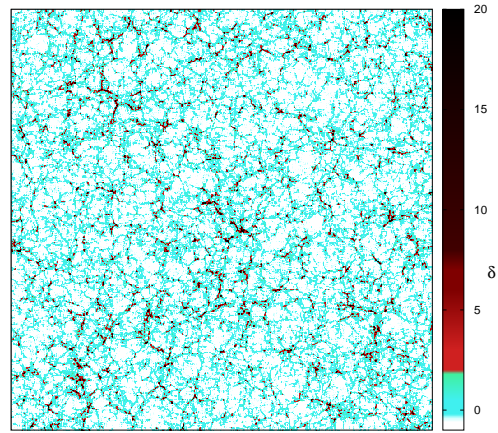}
   \includegraphics[width=0.8\columnwidth]{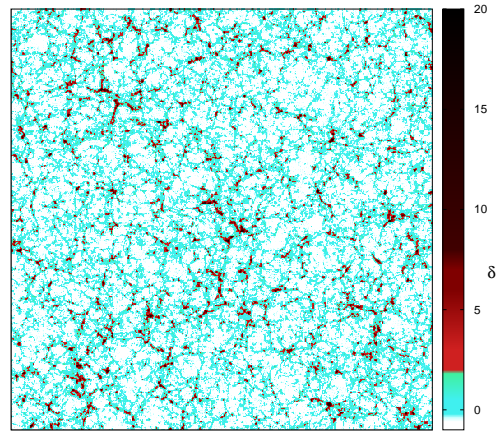}
   \caption{Here we show the difference between performing an actual \nbody\ simulation (top) and using 2LPT (bottom) to generate a particle distribution at $z=0.5$, with the same initial conditions.
   The image shows a slice of the density constrast $\delta$ distribution in a ${1h^{-1}{\rm Gpc}}^3$ box.
   }
   \label{fig:2lpt}
 \end{center}
 \end{figure}

The basic scaffolding of \halogen\ is an appropriate dark matter density field realised at the desired redshift, sampled by $N$ particles. For simplicity we choose to use $2^{\rm nd}$-order Perturbation Theory (2LPT) \citep{2lpt_0,bouchet}
to produce this field,
which can be obtained with the public code \textsc{2LPTic}.

   We show in \Fig{fig:2lpt} the density distribution of an \nbody\ simulation (top panel) and a 2LPT representation (bottom panel) at $z=0.5$. 
Notably, the $\textsc{2LPT}$ distribution appears to be blurred in comparison to the \nbody\ simulation.
   This is due to the fact that $\textsc{2LPTic}$ -- as the name suggests -- was originally designed only to generate initial conditions \citep{2lptcode}, since even 2$^{nd}$-order 
  perturbation theory breaks down at low redshift when over-densities become highly nonlinear. The small-scale difference in \Fig{fig:2lpt} can be explained by
  shell crossing,
  an effect in which particles following their 2LPT trajectories cross paths and continue rather than gravitationally attracting each other in a fully non-linear
  manner \citep{shell-crossing,shell-crossing-2}. 
   In order to compensate for shell-crossing, \citet{PTHalos_MM} advocates the use of a smoothing kernel over the input power spectrum. We tested the effect of this smoothing in \halogen\ but did not find 
   any improvement in the final catalogue.
  
  Nevertheless, 2LPT provides a suitable approximation of the large scale distribution of matter, where perturbations have not yet entered into the highly non-linear 
regime and this is sufficient for \halogen.
  Note that \halogen\ is in principle agnostic about the method in which this density field snapshot is produced. Other methods, for instance
  the ``Quick-PM" 
  \citetext{cf. the QPM method described by \citealt{qpm}}, 
  COLA \citep{cola} 
  or 3LPT could equally be employed by the user.  
  A different choice of density field will yield somewhat different results, especially at smaller scales. As long as the chosen method reconstructs large scales correctly, the remaining steps
  of \halogen\ should be unmodified.

  Despite this, we have by default incorporated $\textsc{2LPTic}$ as part of the \halogen\ code (which bypasses the costly I/O of writing the snapshot to disk), but also allow the user to provide an arbitrary snapshot with a 
  distribution of $N$ particles in a cosmological volume.  
  Our choice for $\textsc{2LPT}$ was mainly driven by its low computational cost and success in the distribution of matter at large scales. We use this approach for 
  all results in this paper.

\subsection{The Mass Function} \label{sec:mf}

The halo mass function (HMF) $n(>M)$ measures the number density of haloes 
above a given mass scale. It is required to generate mass-conditional clustering, which in turn is a pre-requisite for extension to HOD-based galaxy mock generation.

We produce a sampled mass function by the standard inverse-CDF method, utilising an arbitrary input HMF.

The most accurate HMF for a given cosmology, over a range of suitable scales, may be obtained from an \nbody\ simulation via a halo-finding algorithm --although
there are notable variations depending on the technique \citep{Knebe11}. 
Since we require a full \nbody\ simulation for the tuning of \halogen, it would be perfectly acceptable to use this simulation to generate the HMF. 
However, in the hope of future improvements, we wish to avoid using the full simulation as far as possible. 
Fortunately, there is a wealth of literature concerning accurate predictions 
of the HMF for widely varying cosmologies and 
redshifts using Extended Press-Schechter theory \citep{Press1974,EPS}. 

The mass function may be calculated by any means, so long as a discretised function of $n(>M)$ is provided. 
For simplicity, we decided to use the online halo mass function calculator \hmfcalc\footnote{\url{http://hmf.icrar.org}}  \citep{hmfcalc} 
for obtaining the halo mass distribution in this paper.
%
%
%
%
%
%
%

In the remainder of the paper we use the fit of \citet{watson} for \bigmd\, and that of \citet{Tinker08} for \goliat\, 
which both constitute reliable fits.

\subsection{Spatial Placement of Haloes} \label{sec:placement}

The crucial step in the generation of approximate halo catalogues is the commissioning of halo positions. In keeping with the philosophy of modularity,
the halo-placement step is de-coupled from the rest. Any routine which takes a vector of halo masses and an array of dark matter particle positions and returns a 
subset of those positions as the halo locations is acceptable. However, we consider this step to be at the heart of the \halogen\ method, as it is responsible 
of generating the correct mass-dependent clustering. 

To achieve an efficient placement that reconstitutes the target two-point statistics, we recognise the validity of the clustering on large scales from the broad-brush 2LPT field. 
We place haloes on 2LPT field particles, essentially using the estimated density field as scaffolding on which to build an approximate halo field. We will follow a 
series of steps in the construction of the method of spatial placement to be presented in \Sec{sec:halogen} below.

\subsection{Assignment of Velocities} \label{sec:velocities}
The most obvious way to assign velocities to each halo would be to use the velocity of the particle on which it is centred. 
However, haloes are viralised systems whose velocities tend to be lower than that of their constituent particles. 
This is potentially mitigated by using the average velocity of all particles within a defined radius of the artificially placed halo. 
However, this is not robust as there are often very few particles inside the halo radius. 
Additionally, 
the 2LPT particle velocities will differ from their \nbody\ counterparts due to shell-crossing, especially on the small scales associated with haloes. 

Thus, we prefer to take a phenomenological approach, and assume that a simple 
mapping via a factor $f_{\rm vel}$ can be applied to the collection of halo velocities to recover the 
results of the \nbody\ distribution
\begin{equation} \label{eq:vel}
{\bf v}_{\rm halo}=f_{\rm vel}\cdot \bf{v}_{\rm part}.
\end{equation}

This factor could \textit{a priori} depend on the velocity (i.e. a non-linear mapping) and the mass of the halo $f_{\rm vel}(v_{part},M_{\rm halo})$.
However, we will show in \Sec{sec:fvel} that a linear mapping is sufficient and present a way to compute
$f_{\rm vel} (M_{\rm halo})$.

\section{HALOGEN}
\label{sec:halogen}

Though \halogen\ is a four-stage proccess, the most crucial aspect
is the assignment of halo positions, which this section describes in some detail. 
The general concept is to specify a sample of particles from an underlying density field as haloes.

The motivating philosophy of \halogen\ is to start from the simplest idea and improve if necessary.
In this vein, we present here successive stages of evolution of the \halogen\ method, which we hope will show satisfactorily that the method as it stands is optimal.
\Fig{fig:construction} will serve as the showcase for the various stages of \halogen. In it we present the 2-point correlation function (2PCF) for each stage of development
to verify that the method approaches the \goliat\ reference catalogue as new characteristics are added.  

Note that the 2PCF is computed with the publicly available parallel code \textsc{CUTE}\footnote{http://members.ift.uam-csic.es/dmonge/CUTE.html} \citep{CUTE}. In the fitting routine 
that is included in the \halogen\ package and described in \Sec{sec:fit} we also use the same code. 

\subsection{Random particles}
\label{sec:rand_no}

We start with the simplest approach: using \textit{random} particles from the 2LPT snapshot as the sites for haloes. 
We expect to recover the large-scale shape of the 2PCF in this way, as this is encoded in the 2LPT density field which we trace.

However, it is clear from \Fig{fig:construction} that this method ('random no-exc')
consistently underestimates the 2PCF over all scales except $r<1\hMpc$, where it should sharply drop to -1, but rather remains positive. 

The consistent under-estimate is a realisation of an inaccurate linear bias, $b$, defined as the scaling factor between the 2-point function
of the haloes and the underlying matter density field:
\begin{equation} \label{eq:rhalo}
\xi_{\rm halo}(r) = b^{2} \xi_{\rm dm}(r)
\end{equation}
We begin to address this in \Sec{sec:ranked}. 

The small-scale clustering can be explained by the fact that particles can be arbitrarily close, whereas distinct haloes -- recall that subhaloes have been removed -- have a well-defined minimum separation (otherwise they merge). 
The turn-over in the simulation based 2PCF occurs around the mean halo radius scale.

\subsection{Random particles (with exclusion)}
\label{sec:rand}

The simplest improvement to the random case is to eliminate the artificial small-scale correlations. 
Though the primary application of \halogen\ will be for large scales, a simple improvement at small scales is useful. 

As we have noted, the artificial clustering at small scales arises from the fact that particles can be arbitrarily close, whereas simulated haloes have a minimum separation.
The radius of a halo is a rather subjective quantity, and its definition is modified in various applications and halo-finders. However, we may parametrise this by
\begin{equation} \label{eq:rhalo}
R_{\Delta}=\Bigg( \frac{3M_{\rm halo}}{4\pi \Delta_h \rho_{\rm crit}} \Bigg)^{1/3},
\end{equation} 
where $\Delta_h$ is the overdensity of the halo with respect to the critical density of the Universe. 
For the work presented here we used $\Delta_h=200$.

Using this scale, we introduce \textit{exclusion}, a modifiable option which controls the degree to which haloes can overlap, which we set to mimic the halo finder's specification.
For example, in this work we use both \ahf\ and \fof\ \citep[see][for a comparison and an introduction to all relevant halo finding techniques]{Knebe:2013}.
For the latter we do not allow any overlap whereas for the former \halogen's halo centres are not allowed to lie inside another halo's radius.

The effect of \textit{exclusion} is presented in \Fig{fig:construction} ('random exc'). 
As expected, scales of $r<1$\hMpc\ show a turnover  while larger scales are unaffected. 
We note that the turnover is at smaller scales for \halogen\ than for \ahf. This is to be expected, as it is unlikely to find two \ahf\ haloes separated
by a distance slightly exceeding $R_{\Delta}$, due to reasons akin to the FOF overlinking problem. In such cases, there is an increased likelihood of the two halos being subsumed into one, or one becoming a subhalo of the other. 
It is conceivable that one could empirically model these effects by tuning the value of $\Delta_h$ by some factor which captures this suppressed probability.
However, as we are more interested in large scales and these considerations touch upon the subtleties of halo definition, we consider these exclusion criteria sufficient for present purposes. 
We will use this form of exclusion (in an appropriate form) for all following work.

\subsection{Ranked approach} \label{sec:ranked}
We return now to the problem of under-estimation of the correlations, which we noted was due to an incorrect realisation of the linear halo bias. In effect, a random choice of particle position correponds to sampling the matter power spectrum uniformly, and therefore $b=1$. 
However, halo bias is generally greater than unity (especially for higher mass halo samples) \citep{Tinker05}. 

Increasing the bias corresponds to sampling higher-density regions. The simplest way to achieve this is to rank-order the density of regions in the particle distribution, and assign haloes to these regions based on their mass. 

To calculate densities from the particle distribution, we simply create a uniform grid with cell-size $\l_{\rm cell}$, and obtain the density in each cell using a Nearest-Grid-Point (NGP) assignment scheme \citep{Hockney:1988}. We consider specification of the optimal $l_{\rm cell}$ in \Sec{sec:lcell}. The cells are ordered by density, and the haloes by mass, and each halo is assigned to its corresponding cell (a random particle is chosen within the cell). 

Using $l_{\rm cell}=5\hMpc$ in this case, we obtain the results shown in \Fig{fig:construction} labelled  'ranked exc'. The resulting 2PCF is now overestimated. 
This is not surprising, since even if we expect haloes to form in dense environments, the bias is not completely deterministic: 
in reality the $n^{\rm th}$ most massive halo does not need to 
reside in the $n^{\rm th}$ densest place. 

The effect of introducing a scale length, $\l_{\rm cell}$, is also clearly seen in this result. There is a turnover in the 2PCF below $l_{\rm cell}$, which corresponds 
to a significant reduction of bias on these scales since a random particle is chosen within the cell.

\subsection{$\alpha$ approach}
\label{sec:alpha}
We find that selecting completely random particles yields too low a bias, whereas the ranked approach is highly biased. We require an intermediate solution, which has higher probability of selecting dense areas than the random approach, and 
lower probability than the ranked approach.

The probability that a cell is chosen is a function of its density,
\begin{equation} \label{eq:prob}
	P_{\rm cell } \propto G(\rc).
\end{equation}
In the completely random case, we have $G(\rc) = \rc$. 
In principle we can tailor $G(\rc)$ so that the probability of selecting a cell reproduces the appropriate bias.
We choose to constrain $G(\rc)$ to have a power-law form, i.e.
\begin{equation}
\label{eq:alpha}
	G(\rc) = \rc^\alpha.
\end{equation}
When $\alpha = 1$, we recover the random approach, and as $\alpha \to \infty$ we obtain the ranked approach.

\begin{figure}
  \centering
  \includegraphics[height=\linewidth,angle=270]{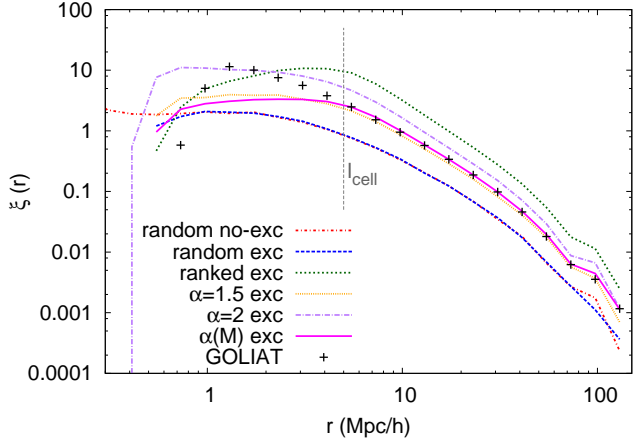}
  \caption{Two-point correlation function of the \goliat\ haloes in comparison to \halogen\ for the various evolutionary stages presented in Sections~\ref{sec:rand_no}
  through \ref{sec:alphaM}. The dashed vertical line indicates the cell size of $l_{\rm cell}=5$\hMpc\ applied for the approaches \ref{sec:ranked} through \ref{sec:alphaM}.}
  \label{fig:construction}
\end{figure}

In \Fig{fig:construction} we show results for $\alpha = 1.5, 2$, 
demonstrating the effectiveness of our model for tuning the normalisation (i.e. bias) of the 2PCF. The $\alpha = 1.5$ curve closely matches the 2PCF of the \ahf\ catalogue,
at least at scales larger than the applied cell size $l_{\rm cell}=5\hMpc$.

The exact value of $\alpha$ for a particular application may be determined by a least-squares fit, which we describe in more detail in \Sec{sec:fit} (note that here the choice of $\alpha$ was not formally fit).

In corollary with this prescription, we also introduce a means to roughly ensure 
\textit{mass conservation} in cells: once a halo is placed, if the total halo mass in the cell exceeds the original mass, the cell is eliminated from future 
selections. However, we do not update the value of the probability after every halo placement because it is computationally very expensive ($O(N_{\rm cell}^3)$) 
and we have checked that doing so has a negligible effect on output statistics.

We note that a similar method was employed in QPM \citep{qpm}. In fact, the physically meaningful quantiy is $f_{\rm halo}(\rho)$ -- the distribution of halo density (i.e. the fraction of haloes in cells with density $\rho$). 
This can be written as 
\begin{equation}
\label{eq:distr}
	f_{\rm halo}(\rho) = P({\rm cell}|\rho) f_{\rm cell}(\rho),
\end{equation}
where $P({\rm cell}|\rho)$ specifies the relative probability of choosing a cell given its density (in our case, $\rho^\alpha$), and $f_{\rm cell}(\rho)$ is the intrinsic distribution of cell densities given the cell size and cosmology (heavily related to the cosmological parameter $\sigma_8$). 
QPM specifies the target distribution $f_{\rm halo}(\rho)$ directly, as a Gaussian. In \halogen\, we instead specify $ P({\rm cell}|\rho)$, which is more closely tied to our algorithm.
In principle one can convert from QPM-like methods to \halogen\ with \Eq{eq:distr}.

\subsection{$\alpha(M)$ approach} \label{sec:alphaM}
The approach as it stands reproduces the 2PCF accurately down to the scale of $l_{\rm cell}$. If the 2PCF of a sample of given number density is all that is required for a specific application, then this will do well. 

However, if we were to select a sub-sample of the most massive haloes of our catalogues and recompute the 2PCF, the bias would be incorrect, since more massive haloes are more biased
\citep{Tinker05}. For a truly representative catalogue, in which the haloes are conditionally placed based on their mass, the bias model is required to be mass-dependent.
Failing this, there is no physical meaning attached to the assignment of masses in the second step (\Sec{sec:mf}). 

Mass-dependent halo bias is also crucial for implementing HOD models on the catalogue, for use in galaxy survey statistics,
as the number of galaxies associated with a halo depends on its mass.

We incorporate this mass-dependence into the $\alpha$ parameter, so that we finally have
\begin{equation}
	G(\rc,M) = \rc^{\alpha(M)},
\end{equation}
with $\alpha (M)$ an increasing function.

In practice, we use discrete mass bins, and for each bin $i$, with masses $M^{i-1}_{\rm th}>M>M^i_{\rm th}$, we use a different $\alpha_i$. We describe how we obtain the 
best-fit
to this mass-dependent $\alpha$ using the fiducial halo catalogue from the simulation in \Sec{sec:fit}.

Using just five mass bins, we illustrate this approach in \Fig{fig:construction}, labelled ``$\alpha(M)$ exc" (magenta line) using the best-fit values for $\alpha(M)$.
We list in \Tab{tab:bins_goliat} the mass
thresholds, applied $\alpha$-values, and corresponding number densities of all haloes with $M_{\rm halo}>M_{\rm th}^i$. 
Note that though the probability is not recomputed after placing a halo, it is recomputed with updated $\rho$ and $\alpha$ when changing mass bins. 

Though the $\alpha(M)$ approach does not improve the 2PCF with respect to the $\alpha$ approach in \Fig{fig:construction}, it has the clear advantage of reproducing a mass dependent clustering, which as we noted is essential for further HOD analyses, and useful for being able to use any mass-range in the same realisation.

  \begin{table}
  \begin{center}
  
      \begin{tabular}[width=\linewidth]{cccc}
      \hline
       bin 	& $M_{{\rm th}}^{i}$ [\hMsun]& $n_i$ [(\hMpc$)^{-3}$]	& $\alpha_i$ \\
      \hline
	0 		& $1.64\cdot10^{14}$	& $0.05\cdot10^{-4}$ 	& 3.54	\\
	1 		& $4.80\cdot10^{13}$	& $0.40\cdot10^{-4}$ 	& 2.26	\\
	2 		& $2.65\cdot10^{13}$	& $0.90\cdot10^{-4}$ 	& 1.77	\\
	3 		& $1.86\cdot10^{13}$	& $1.40\cdot10^{-4}$ 	& 1.48	\\
	4 	 	& $1.38\cdot10^{13}$	& $2.00\cdot10^{-4}$	& 1.41	\\
      \hline
      \end{tabular}
      \caption{Properties of the selected mass bins for the \goliat\ simulation: mass threshold $M^i_{\rm th}$, equivalent number density $n(M>M_{\rm th}^i)$ and best fit
      $\alpha_i$ in $M_{th}^{i-1}<M<M_{th}^{i}$ for the \halogen\ 
      $\alpha(M)$ approach.} 
  \label{tab:bins_goliat}
  
  \end{center}
\end{table}

\subsection{Summary}
In conclusion, \halogen\ constitutes a method for generating a halo catalogue which exhibits correct 2-point clustering statistics, while not only positioning the haloes correctly, but also imbuing them with physically meaningful masses. 
The method can be summarised as follows.

The particles generated by 2LPT (\Sec{sec:2lpt}) are covered by a  grid of cell size $l_{\rm cell}$, the halo masses $M_i$ generated from the halo mass function (\Sec{sec:mf}) are ordered by mass, and starting from the most massive halo
they are placed by

	\begin{enumerate}
		\item selecting a cell with probability $P_{\rm cell} \propto \rho_{\rm cell} ^{\alpha(M)}$,
		\item randomly selecting a particle within the cell and using its coordinates as the halo position,
		\item ensuring that the halo does not overlap (following an \textit{exclusion} criterion) with any previously placed halo in any cell, and re-choosing a different 
		random particle in that case,\footnote{If, after several iterations all the particles found were excluded, re-choose cell (to avoid infinite loops).}
		\item subtracting the halo's mass from the selected cell, $m_{\rm cell} = m_{\rm cell} - M$: if $m_{\rm cell} \leq 0$ the cell is removed from selection.
	\end{enumerate}
	
Note that the physically motivated nature of the process suggests that higher-order statistics may also be recovered with some success. 

\section{Parameter study}
\label{sec:parameters}
We have mentioned several parameters of the \halogen\ method, and these are of particular importance in producing accurate realisations. 
In this section we will discuss each parameter, its effects and how to optimise for it if possible. 

There are three parameters in \halogen\ (with other options and parameters being expressly determined by the required output, such as the size of the simulation box $L$): 
the two physical parameters of the model, $\alpha$ -- controlling the linear bias -- and $f_{\rm vel}$ -- controlling the velocity bias -- and the one parameter of the algorithm, $\l_{\rm cell}$.  
These are summarised in Table \ref{tab:parameters}, and detailed in the following subsections.

In the previous Section we used \goliat\ as a reference. We now turn to \bigmd\ and its 
\fof\ catalogue: this simulation has a larger volume, allowing us to probe BAO scales. The increased volume also reduces cosmic variance on intermediate scales.
\halogen\ primarily aims at reproducing clustering statistics for even larger volumes, 
hence it is beneficial to assess the performance of \halogen\ and its parameters in this regime. Furthermore, this demonstrates independence
from the underlying simulation and halo finding technique.

  \begin{table}
  \begin{center}
  
      \begin{tabular}{cccc}
      \hline
      \textsc{Parameter} &  \textsc{Motivation} & \textsc{Value}  \\
      \hline
      $\alpha_i$		& linear bias & $\chi^2$-fit to bias 		\\
      $f^i_{\rm vel}$  &	velocity bias & $f^i_{\rm vel}=\sigma^i_{\rm NB}/\sigma^i_{\rm p}$		\\
      $l_{\rm cell}$ &	algorithm	& $l_{\rm cell} \approx 2\cdot d_{\rm p}$ \\
      
      \hline
      \end{tabular}
      \caption{A summary of the parameters involved in \halogen, the motivation to introduce them and how to compute/optimise them. See text for details} 
  \label{tab:parameters}

  \end{center}
\end{table}

\subsection{Fitting $\boldsymbol{\alpha (M)}$} \label{sec:fit}

The value of $\alpha(M)$ is crucial to the performance of \halogen, as it constitutes the only physical parameter controlling the bias. 
The \halogen\ package contains a stand-alone routine which determines a best-fit for $\alpha(M)$, which can then be passed to \halogen\ to generate 
any number of realisations. 
We describe this routine here, and illustrate it with application to \bigmd. 
The fitting of $\alpha(M)$ is based on the standard $\chi^2$-minimisation technique. However, a few details are worth mentioning.

\paragraph*{Mass-dependence.}
We perform the fit in sharp-edged mass bins to determine a mass-dependent $\alpha(M)$, i.e. for each bin $i$ we fit a $\alpha_i$ for the mass range $M_{\rm th}^{i-1}<M<M_{\rm th}^i$.
There are two conceivable ways of doing this -- differentially or cumulatively. We have experimented with both and find that the cumulative procedure has better
performance. That is, we fit the first mass bin, and then the first and second together (keeping the best value of $\alpha_0$ for the first bin), and so on. 
This has the advantage of being able to properly correct for deviations in previous bins, which is particularly important since the first bins to be fit are the high masses, for which fewer haloes exist. Misestimation of $\alpha$ here is more likely, but is compensated for when fitting to lower mass bins by including the high-mass estimates in the fit.

\paragraph*{HALOGEN variance.}
The halo placement in \halogen\ is probabilistic, even given a constant underlying density field. Using different random seeds can slightly affect the final placement, and
thus the clustering statistics (the extent of this is dependent on the volume, $n$ and $\alpha$). We term this ``\halogen\ variance", and note that it is not to be 
confused with cosmic variance.
 Cosmic variance is introduced by modifying the the random seed of \textsc{2LPTic}, which in effect results in a different realisation of the universe\footnote{Cosmic variance -- strictly speaking -- requires the study of the same volume, but in a different place in the universe. This approach is more
appropriately called 'sampling variance' yet nevertheless the generally accepted technique for generating covariance matrices.}.
During the fit each mass bin is realised several times (ten in the case of \bigmd) with \halogen\ 
to average out the effects of \halogen\ variance, and also provide an 
error $\sigma_{\rm H}$ (computed as the standard deviation) to use in the definition of $\chi^2$.

\paragraph*{$\boldsymbol{\chi^2}$ minimisation.}
The fit is performed by minimising $\chi^2$:
\begin{equation} \label{eq:chi2} 
 \chi^2(\alpha)= \sum_{j}   \Big(      \frac{ \xi_{\rm H}(r_j|\alpha)-\xi_{\rm NB}(r_j) }{ \sigma_{\rm H}(r_j|\alpha) }     \Big)^2
\end{equation}
where  $\xi_{\rm H}$ and $\xi_{\rm NB}$ are the 2PCFs of \halogen\ and the reference catalogue, respectively. 
We note that minimising this statistic is susceptible to systematic errors in \halogen\ in bins where the stochastic error ($\sigma_H$) is much smaller than the systematic error ($\Delta\xi$). This is especially likely when the region of the fit approaches $l_{\rm cell}$. To test whether the region is stable, we may choose a distance estimator to be minimized that treats all scales with the same weight, eg.  $\Delta=(\xi_{\rm H}-\xi_{\rm NB})^2/\xi_{\rm NB}^2$. We have tried with both quantities in our fitted range, and the results are left unchanged, indicating that the range of the fit is stable.

We use a grid of $\alpha$ to cover the expected result for each mass bin. 
We use a cubic spline interpolation over $\chi^2(\alpha)$ to locate a precise minimum for the best-fit $\alpha$.

\paragraph*{Fitting Range.}
We restrict the range of the fit to scales in which the shape of $\xi_{\rm H}(r)/\xi_{\rm NB}(r)$ is flat. 
This corresponds to mid-range scales of
$15 \hMpc< r < 47 \hMpc$, which avoids small-scale effects of \halogen, and large-scale cosmic variance.

\paragraph*{Number of mass bins.}
The number of bins to use in this procedure will depend on the needs of the user, and the size and resolution of the reference simulation. 
It determines the reliability of the mass-dependent clustering.
For \bigmd\ we distribute the haloes into 8 roughly equi-numbered bins with the mass thresholds $M_{\rm{th}}^i$ as shown in \Tab{tab:bins}.
In that table we also show the best-fit
$\alpha_i$, and the equivalent number density $n_i$ for each mass threshold.

\begin{figure}
  \centering
  \includegraphics[height=\linewidth,angle=270]{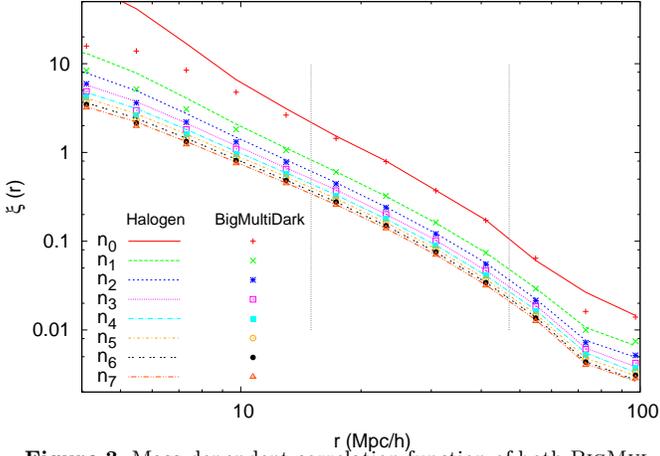}
  \caption{Mass-dependent correlation function of both \bigmd\ (points) and \halogen\ haloes (lines). 
  We select 8 number densities $n_{i}$ (colours in the legend) of haloes, with values found in \Tab{tab:bins} 
  (together with the equivalent mass threshold $M_{\rm th}^i$).
  The vertical dashed lines indicate the range of the fit.}
  \label{fig:MassBins}
\end{figure}

The 2PCFs for our 8 values of $n_i$ 
are shown in \Fig{fig:MassBins}, where we compare the results from \halogen\ against the \bigmd\ reference catalogue. 
The range used during the fitting procedure and for the 
$\chi^2$-minimization is indicated by the vertical lines.

\begin{figure*}
  \centering
  \includegraphics[height=0.48\linewidth,angle=270]{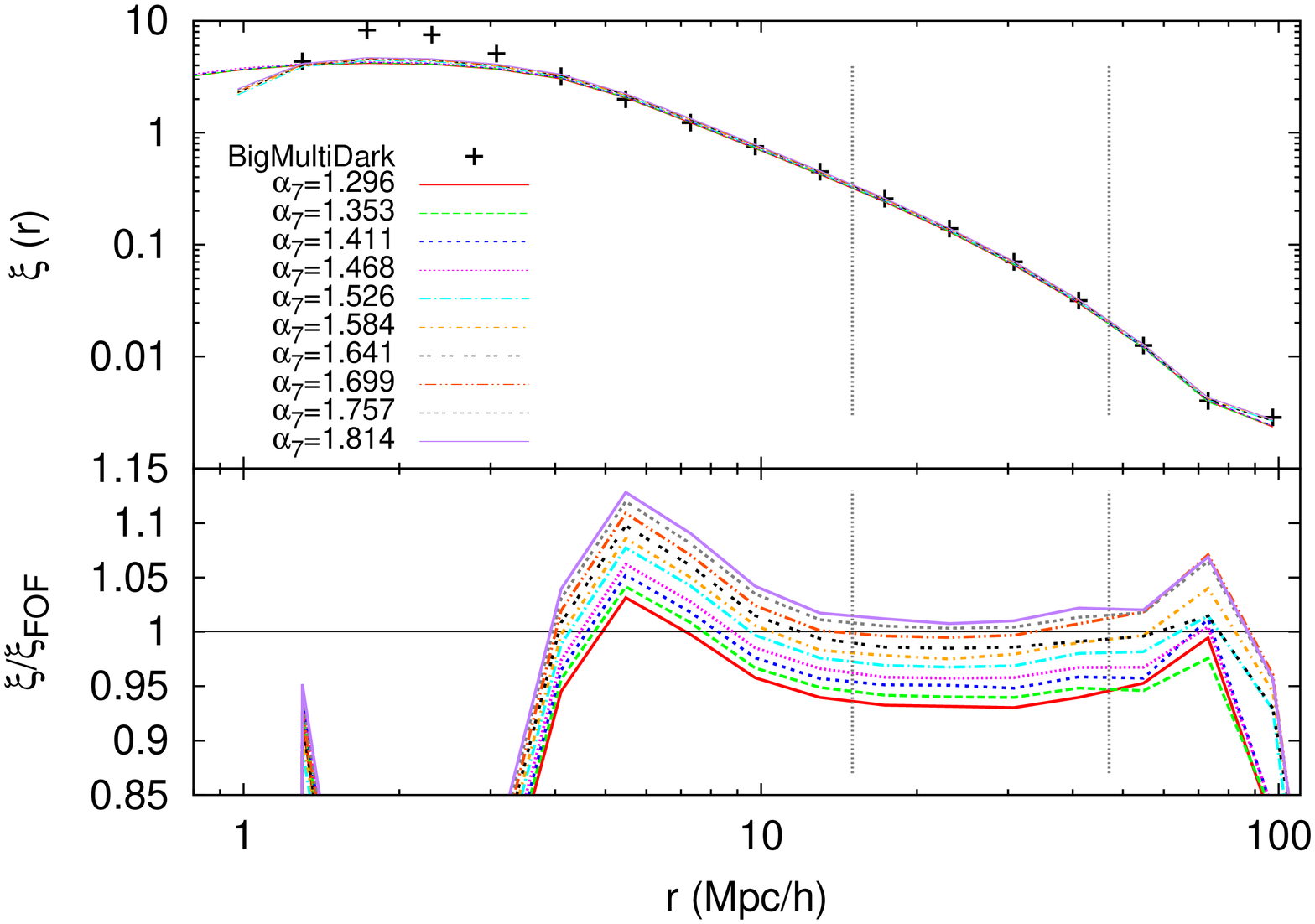}
  \includegraphics[height=0.48\linewidth,angle=270]{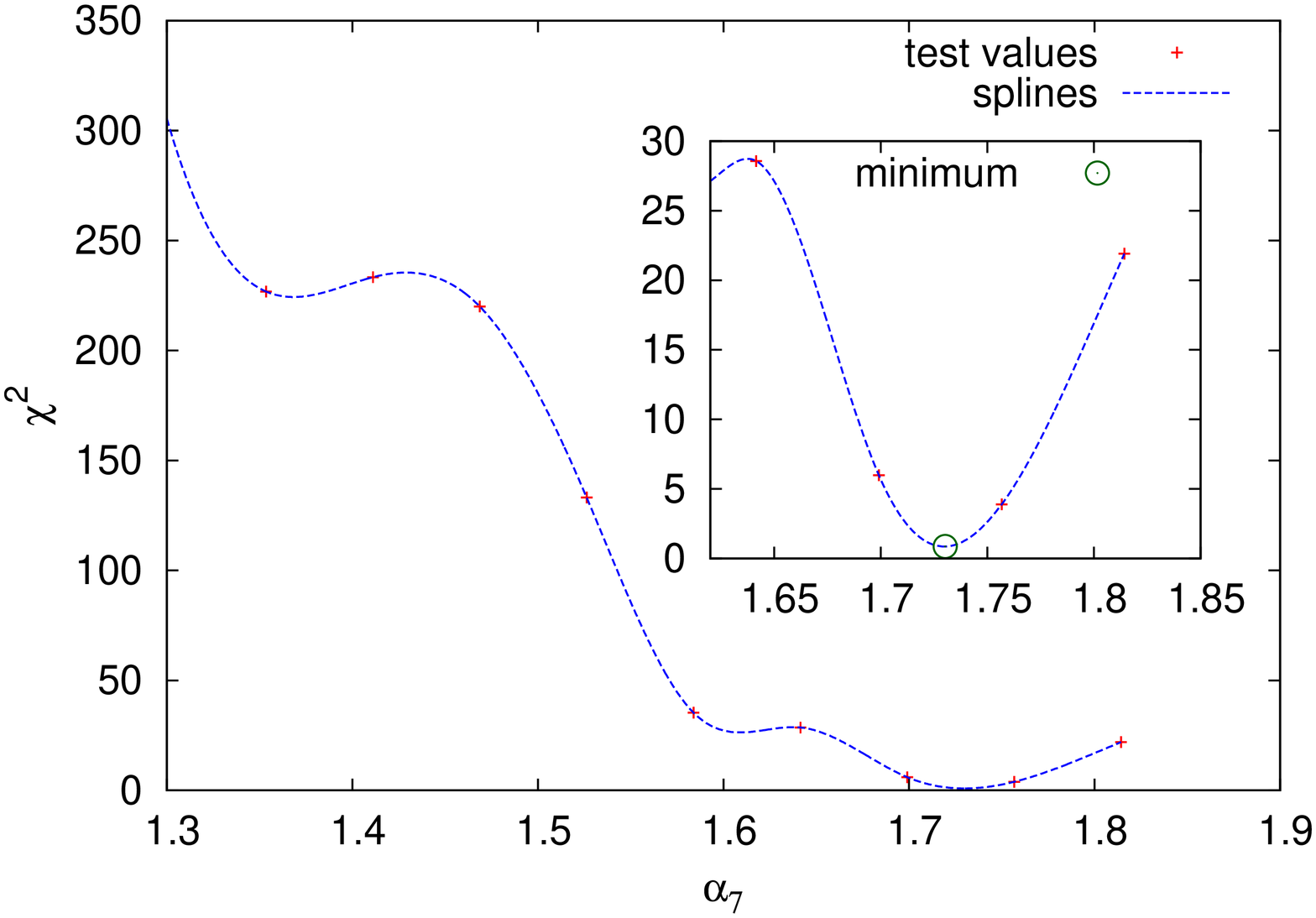}
  \caption{Illustration of variations in $\alpha$ and its consquences for the 2PCF. 
  Left: Correlation function 
  of the target halo catalogue (\bigmd, crosses) and the grid of $\xi_{\rm H}$ corresponding to the grid of $\alpha_7$ used for minimisation. 
  The lower panel shows the ratios to the \bigmd\ result. The vertical dashed lines mark the spatial $r$-range of the fit. 
  Right: $\chi^2$ (\Eq{eq:chi2}) as a function of $\alpha_7$ for the grid of values used in the left panel (red crosses) and the interpolated curve (dashed blue line). 
  In the inner box we zoom into the area near the minimum (green circle). 
  }
  \label{fig:FitAlpha}
\end{figure*}

We note that the choice of $\alpha$ finely controls the bias. This is demonstrated in \Fig{fig:FitAlpha}, in which we show the resultant $\xi(r)$ for the entire grid of 
$\alpha_7$ for this fit (left-hand side). There is a $\sim 10$ per cent deviation in $\xi_{\rm H}(r)$ over the grid range ($1\%$ between consecutive lines). 
On the right-hand side of the figure, we show the $\chi^2$ of each of those curves and the cubic spline fit interpolation used to find the minimum, which corresponds
to the $\alpha_7$ best-fit value shown in \Tab{tab:bins}.

  \begin{table}
  \centering
      \begin{tabular}{ccccc}
      \hline
       bin $i$ 	& $M_{{\rm th}}^{i}$ [\hMsun]	& $n_i$ [(\hMpc$)^{-3}$]	& $\alpha_i$ & 	$f_{\rm vel}$\\
      \hline
	0	& $1.64\cdot10^{14}$ 	& $0.05\cdot10^{-4}$ 	& 4.80	& 0.564 \\
	1	& $4.93\cdot10^{13}$ 	& $0.45\cdot10^{-4}$ 	& 2.79	& 0.672 \\
	2	& $2.95\cdot10^{13}$ 	& $0.95\cdot10^{-4}$ 	& 2.28	& 0.715 \\
	3	& $2.15\cdot10^{13}$ 	& $1.45\cdot10^{-4}$ 	& 2.00	& 0.743 \\
	4	& $1.70\cdot10^{13}$ 	& $1.95\cdot10^{-4}$ 	& 1.90	& 0.754 \\
	5	& $1.41\cdot10^{13}$ 	& $2.45\cdot10^{-4}$ 	& 1.84	& 0.760 \\
	6	& $1.21\cdot10^{13}$ 	& $2.95\cdot10^{-4}$ 	& 1.73	& 0.771 \\
	7	& $1.04\cdot10^{13}$ 	& $3.50\cdot10^{-4}$ 	& 1.73	& 0.771 \\
      \hline
      \end{tabular}
      \caption{Properties of the selected mass bins for the \bigmd\ simulation: mass threshold $M_{{\rm th}}^{i}$, equivalent number density $n(M>M_{\rm th}^i)$,
      best fit $\alpha_i$ for the interval of masses $M_{th}^{i-1}<M<M_{th}^{i}$ and $f_{vel}$ computed for the same interval (see \Sec{sec:fvel}).}
  \label{tab:bins}
\end{table}

\subsection{Velocity factor $f_{\rm vel}$}
\label{sec:fvel}

In \Sec{sec:velocities} we outlined a method of converting the velocity of \textsc{2LPTic} particles (designated as halo sites), $\bf{v}_{\rm p}$, to the velocity of a \halogen\ halo, $\bf{v}_{\rm h}$. We stated that 
the transformation was linear in $\bf{v}_{\rm p}$, and thus we can write
\begin{equation}
\label{eq:velbias}
	\mathbf{v}_{\rm h} = f_{\rm vel}(M) \cdot \mathbf{v}_{\rm p},
\end{equation}
where we have retained a mass-dependence in the conversion factor. This section will explore the means to calculate this factor.

\begin{figure}
  \centering
  \includegraphics[height=\linewidth,angle=270]{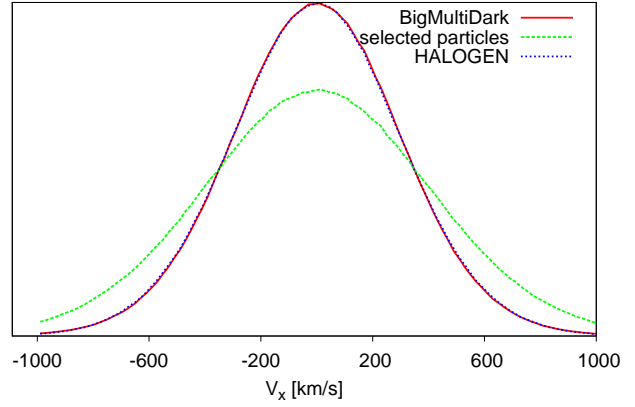}
  \caption{One-component ($v_{x}$) velocity distribution of the halo catalogues. The \fof\ haloes from the \bigmd\ simulation are in a red solid line, and 
  $v_{x,\mathrm{p}}$ of the particles selected by \halogen\ catalogue are in a green dashed line, while the corrected $\bf{v}_{\rm h}$ haloes from \halogen\ are 
  in a blue dotted line. The correction provides a very 
  closely matching distribution, which has a generally lower velocity.}
  \label{fig:vel}
\end{figure}

We begin by justifying our choice of a linear function. 
\Fig{fig:vel} shows the one-component velocity distribution of \bigmd\ and the particles selected by \halogen.
Both curves are well-described by a Gaussian with $\bar{v}_x = 0$, where the standard deviation of the \nbody\ haloes is reduced compared to that of
$v_{x,\mathrm{p}}$, i.e. $\sigma_{\rm p}>\sigma_{\rm NB}$. 
This confirms our claim in \Sec{sec:velocities} that the particle velocities are larger than the 
halo velocities, and also shows that a simple linear transformation suffices to map the distribution of  $\bf{v}_{\rm p} \to \bf{v}_{\rm h}$.

This simple characterisation leads to a transformation of  $f_{\rm vel}=\sigma_{\rm NB}/\sigma_{\rm p}$, which is verified by the blue dotted line where 
this remapping has been applied. 

We expect that the velocity bias \citep{colin2000} will be dependent on mass-scale in general. We can easily incorporate this into our fitting routing by calculating
\begin{equation}
\label{eq:fvel}
 f_{\rm vel}^i=\frac{\sigma^{i}_{\rm NB}}{\sigma^{i}_{\rm p}}
\end{equation}
for each interval of mass $M=(M_{\rm th}^{i-1}:M_{\rm th}^i]$ while performing the fit for $\alpha$. These results are also listed in \Tab{tab:bins}.
There is a noticeable decrease 
in $f_{\rm vel}$ towards higher mass haloes. We will see in \Sec{sec:RSD} below how this affects the modelling 
of Redshift Space Distortions. 

We finally note that there may be other more complex models of velocity bias accounting for the physics of low scales and adjusting other statistics beyond the overall velocity distribution. 
However, the model presented here is very simple and capable of reproducing the halo velocity distribution with a great accuracy.

\begin{figure*}
  \centering
  \includegraphics[height=0.48\linewidth,angle=270]{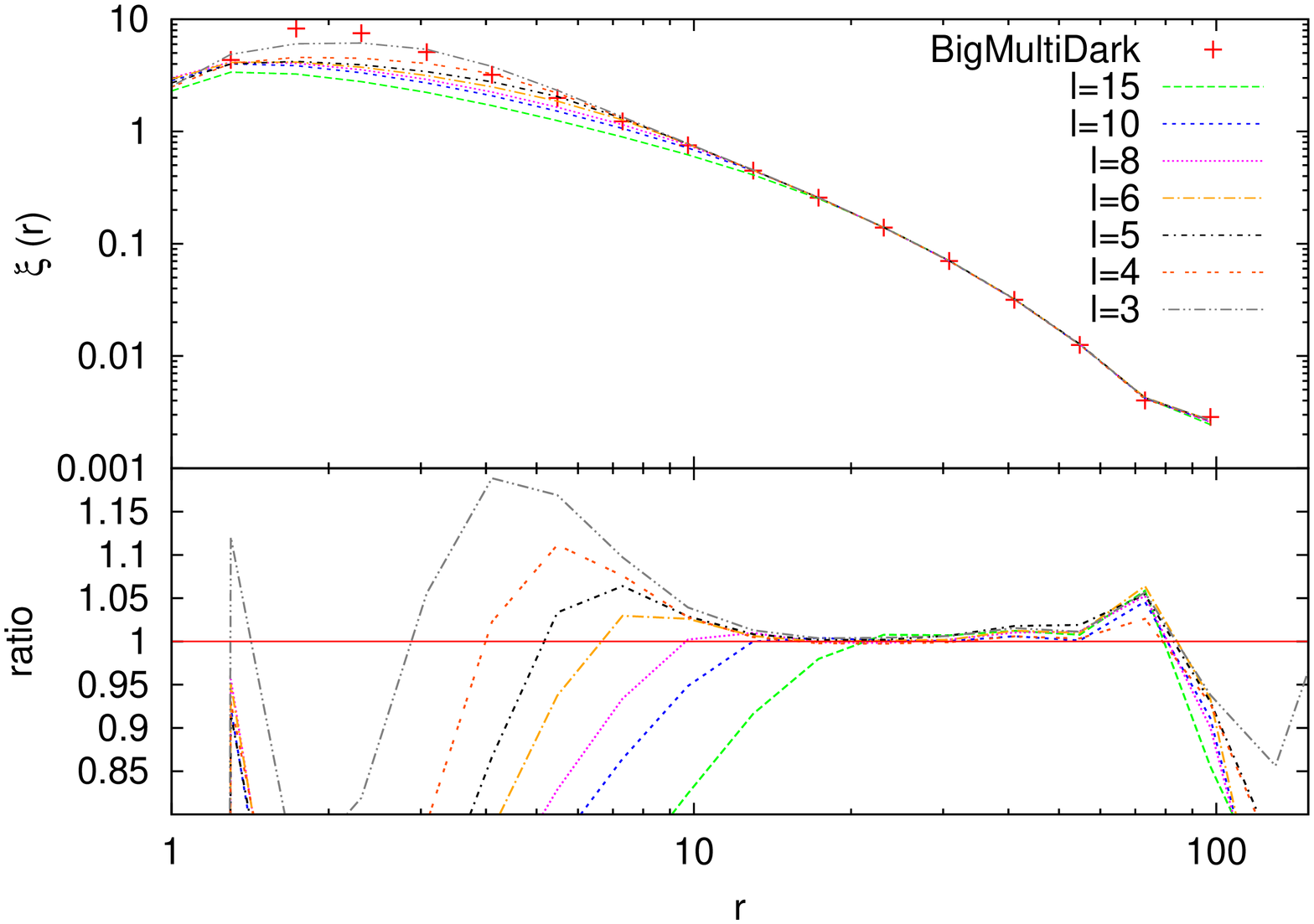}
  \includegraphics[height=0.48\linewidth,angle=270]{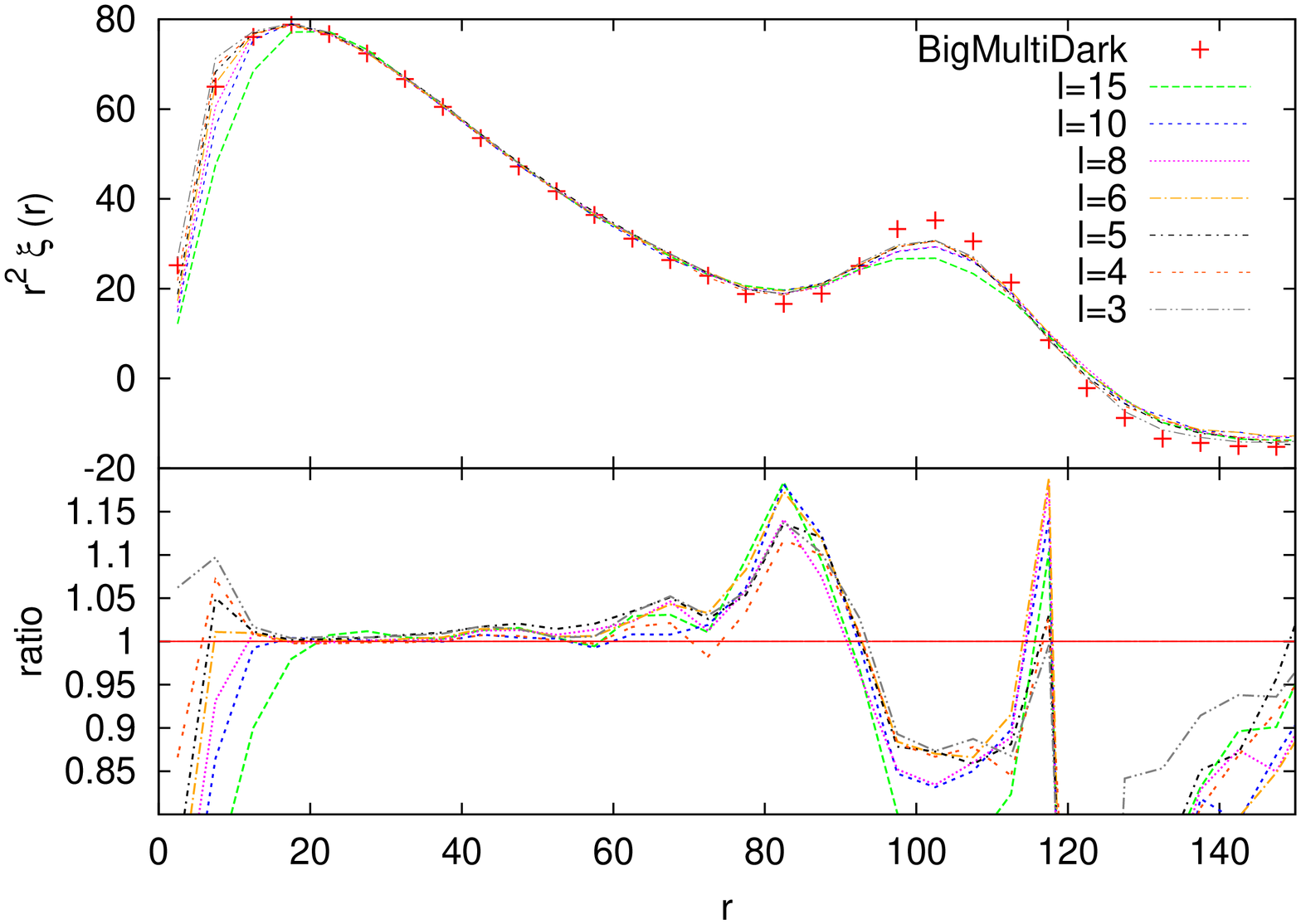}
  \caption{Two-point correlation function on logarithmic (left) and linear (right) scale of the \fof\ catalogue of the \bigmd\ simulation (crosses) against the results from \halogen\ (lines) for different
  values of $l_{\rm cell}$ (different linestyles as indicated in the legend). Note that in the right panel the 2PCF has been multiplied by $r^2$ to increase the visibility of the BAO peak.
  The lower panels show the ratio with respect to the \bigmd\ curve.}
  \label{fig:lcell}
\end{figure*}

\subsection{Cell size: $l_{\rm cell}$} \label{sec:lcell}
We have previously mentioned the cell-size $\l_{\rm cell}$ which is introduced to \halogen\ to provide a simple local density via the NGP scheme \citep{Hockney:1988}. We have also noted that it defines a lower-limit of reliability of the resultant 2PCF. In this section we explore this parameter further, describing its effects and how to optimise for it.

In \Fig{fig:lcell} we show the 2PCF of the \bigmd\ catalogue against \halogen\ results for several values of $\l_{\rm cell}$.
We note two effects, $l_{\rm cell}$ 
\begin{enumerate}
	\item determines the minimum scale at which the 2PCF is reliable and
	\item controls the broadening of the Baryonic Acoustic Oscillations (BAO) peak.
\end{enumerate}
The first effect is clearly noticeable in the left-hand panel where the \halogen\ 2PCF detaches from the \bigmd\ curve at $r\approx l_{\rm cell}$. 
This is expected, since particles are chosen at random inside the cell, reducing the bias at these scales.

The second effect is more noticeable in the right-hand panel. As $\l_{\rm cell}$ is decreased, the broadening and dampening (best seen in the lower panel as the difference between the artificial peak at $r=80\hMpc$ and trough at $r=100\hMpc$) is decreased. The reason for this is that we introduce an uncertainty (on a scale $l_{\rm cell}$) in the position of the haloes that propagates to an uncertainty in the determination of $r_{\rm BAO}$. In effect, the density field has been filtered by a quasi-top-hat function \citep{Hockney:1988}, which has the known effect of peak-broadening.

Clearly, $l_{\rm cell}$ should be set as small as possible to mitigate these effects.
However, 
a limit is enforced by the mean-interparticle-separation, $d_{\rm p}$, of the input density field. We cannot hope to reliably probe scales smaller than $d_{\rm p}$, and even just above this scale we run into the problem of having poor statistics within cells. We recommend using a value of $l_{\rm cell} \geq 1.5d_{\rm p}$ (ensuring $>3$ particles per cell on average), and in this work we take $l_{\rm cell} = 4\hMpc \approx 2d_{\rm p}$ as the reference. 

We comment here that the choice of $l_{\rm cell}$ affects the optimal $\alpha(M)$ relation. This is unfortunate, because it would be useful to be able to perform the fit for $\alpha$ using a lower resolution (since this is the bottleneck). The mechanism by which this effect occurs is known, and we hope to be able to correct for it in the future.

Let us illustrate the mechanism with an example: suppose we take a cell with cell-size $l^{\rm I}_{\rm cell}$ and density $\rho^{\rm I}_{\rm cell}$ from a volume $(Nl^{\rm I}_{\rm cell})^3$.
For the same distribution, we could also use $l^{\rm II}_{\rm cell}=l^{\rm I}_{\rm cell}/2$, which forms 8 sub-cells $i$ with densities $\rho^{\rm II}_{{\rm cell},i}$. For the same $\alpha$,
the probability of choosing the cell in case I is
\begin{equation}\label{eq:caseI}
	P^{\rm I}_{\rm cell} = \frac{(\rho^{\rm I}_{\rm cell})^{\alpha}}{\sum_j^{N^3} (\rho^{\rm I}_j)^\alpha} = \frac{\left(\frac{1}{8}\sum_i^8 \rho^{\rm II}_{{\rm cell},i}\right)^\alpha}{\sum_j^{N^3} (\rho^{\rm I}_j)^\alpha}
\end{equation}
whereas in case II we have
\begin{equation}\label{eq:caseII}
	P^{\rm II}_{\rm cell} = \frac{\sum_i^8 (\rho^{\rm II}_{{\rm cell},i})^\alpha}{\sum_j^{(2N)^3} (\rho^{\rm II}_j)^\alpha},
\end{equation}
and clearly these are not in general equivalent if $\alpha \neq 1$. 
We expect the difference in the distributions to be dependent on $\alpha$, the two cell-sizes and their ratio and the cosmology, via the mass variance $\sigma(r)$. In future studies we hope to be able to quantify this relationship to enable faster fitting.

\begin{figure}
  \centering
  \includegraphics[height=\linewidth,angle=270]{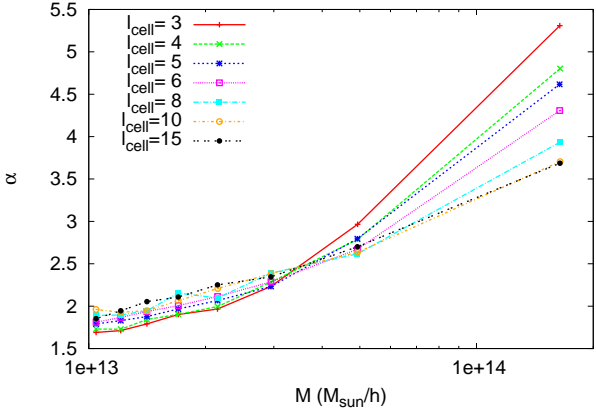}
  \caption{Best-fit $\alpha(M)$ functions for different values of $l_{\rm cell}$, as marked in the legend (units of $\hMpc$).}
  \label{fig:Malpha}
\end{figure}

\Fig{fig:Malpha} shows the effect of changing $l_{\rm cell}$ on the best-fit $\alpha(M)$ and we notice two characteristics. 
Firstly, $\alpha(M)$ is an increasing function for all $l_{\rm cell}$, as expected since $b(M)$ is increasing.
Secondly, low masses are less sensitive to $l_{\rm cell}$, which we expect mathematically from Eqs.\ref{eq:caseI} and \ref{eq:caseII} with an increasing $\alpha(M)$ 
(the greater $\alpha$ is, the greater the differences expected).

In \Fig{fig:lcell} we have re-fit the $\alpha(M)$ relation for each value of $l_{\rm cell}$, ensuring proper comparison between curves. Furthermore, we run 5 realisations of each and display the average, to reduce the effects of \halogen\ variance.

\section{Results and Applications}\label{sec:applications}

While previous sections were dedicated to the dessign and optimisation of \halogen, we have now defined the final method and fixed the optimal parameters.
In this section we discuss the performance of \halogen\ in more detail, both in the clustering statistics so far analysed, and in other statistics 
that \halogen\ is not constrained to match. We begin by demonstrating the power of \halogen\ for mass-production of halo catalogues for use in deriving 
covariance matrices to measure cosmic 
variance, which we envision as the primary application of the \halogen\ machinery.

\subsection{Mass production of halo catalogues}
\label{sec:new}
The driving motivation of developing fast methods for synthetic halo catalogues is to accurately produce robust covariance matrices for large galaxy survey statistics. 
Though \halogen\ requires a full \nbody\ simulation to calibrate its two parameter sets, once these parameters have been established, we are free to run as many realisations 
(with different phases for the initial conditions) of the the halo catalogue (using the same cosmological parameters, volume, mass resolution etc.) as we like. 
This process is expected to purely simulate the effects of cosmic variance, and thus is extremely valuable for deriving the covariance matrices.

In order to verify that the variance seen in the resulting data traces the expected cosmic variance, we complemented the generation of the \halogen\ catalogues with several corresponding \nbody\ simulations. 
Due to the computational time constraints, we were only able to run five simulations, which were based on \goliat, and in which only the seed for
the random Initial Condition (IC) phases was changed. The initial conditions for these runs were generated with \textsc{2LPTic} at redshift $z=32$ (for the \nbody) and $z=0$ (for \halogen), 
using the same seed for each pair. The \nbody\ particle distributions were evolved to $z=0$ using \gadget2 (and subsequently analysed with \ahf).

\begin{figure}
  \centering
  \includegraphics[height=\linewidth,angle=270]{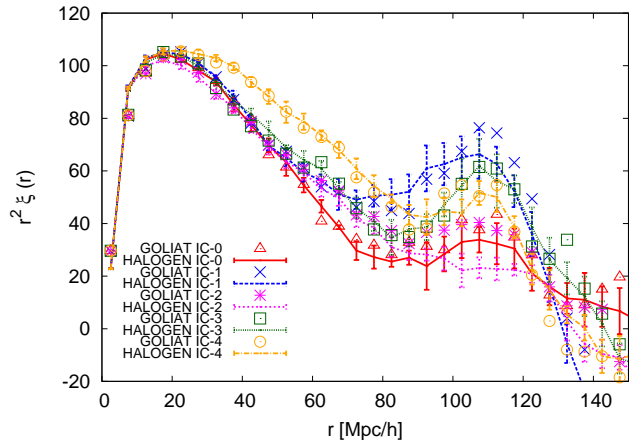}
  \caption{2PCF of \halogen\ (lines) and the AHF (points) catalogues for five different \textsc{2LPTic} random seeds. The first case corresponds to the 
  original \goliat\ used to obtain the $\alpha(M)$ relation
  whereas the following share the same setup besides the seed. All the \halogen\ lines have been averaged over five realisations 
  and the error bars show the \halogen\ variance. Similarity in goodness of fit between the first case and the others indicates that the fitted $\alpha(M)$ is universal with respect to input seed.
  }
  \label{fig:seed}
\end{figure}

In \Fig{fig:seed} we present the 2PCF of those 5 pairs of catalogues (with \halogen\ as lines, and \ahf\ as points). 
The \halogen\ lines are the average of 5 realisations of \halogen\ placement (maintaining the same phases) and the error bars show the \halogen\ variance.
Given that the \goliat\ box size is rather small ($1h^{-1}$Gpc), scales $r>\sim 60 \hMpc$ are dominated by cosmic variance effects. This makes it easy to identify the signature 
of each set of initial conditions. 
Though the realisations are significantly different, we note that the \halogen\ catalogue follows the \nbody\ result, and maintains the correct normalisation at
intermediate scales ($20\hMpc<r<50\hMpc$). We stress that the fitting procedure has only been performed once; all five cases used fixed parameters. The similarity of the goodness of fit
in each case (as compared to that directly fitted to) demonstrates that the fitted $\alpha(M)$ is universal with respect to input seed. 
We note also that the \halogen\ variance is significantly subdominant to the cosmic variance.

\begin{figure}
  \centering
  \includegraphics[height=\linewidth,angle=270]{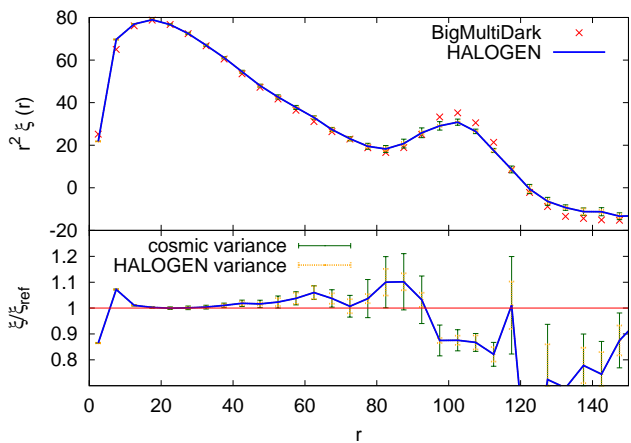}
  \caption{2PCF of the \fof\ catalogue of the \bigmd\ simulation compared to that of a single-run \halogen\ (non-averaged) with the same initial condition phases.
  We also include error bars: in green (solid line) the cosmic variance and in orange (dotted line) the \halogen\ variance (see text). The lower pannel shows the ratio with respect to \bigmd.
  }
  \label{fig:variance}
\end{figure}

To better appreciate the dominance of the cosmic variance in a more applicable 
scenario, we return to the \bigmd\ simulation. 
This has a reduced cosmic variance due to the larger volume,
but has the disadvantage that we cannot run several \nbody\ simulations of this magnitude.
The blue line of \Fig{fig:variance} shows how the 2PCF of a single-run \halogen\ (neither \halogen\ nor cosmic variance  has been averaged out) 
compares to the reference \bigmd\ catalogue when they have the same initial condition phases.
We further show the \halogen\ variance  ($\sigma_{\rm H}$) and cosmic variance ($\sigma_{\rm cosm}$). The former has
been computed as usual: running 5 realisations of \halogen\
on the same 2LPT snapshot. 
For the latter we run five \textsc{2LPTic} snapshots with different IC seeds. In order to avoid mixing $\sigma_{\rm cosm}$ and $\sigma_{\rm H}$
for each of them we first averaged out \halogen\ variance by running 5 realisations of \halogen\, and $\sigma_{\rm cosm}$ is computed as the dispersion of 
the five resulting ($\sigma_{\rm H}$-free) lines. 
We find for all scales that the \halogen\ variance is dominated by the cosmic variance, $\sigma_H<\sigma_{\rm cosm}$.

\subsection{Probability Distribution Function}
\label{sec:PDF}

A simple but powerful statistic for point particles is the Probability Distribution Function (PDF), which is the distribution of particles per cell on a given scale. 
Though simple, it contains interesting information as it contains contributions from the entire hierarchy of $n$-point functions \citep{CIC,CIC-2,CIC-3}.

\begin{figure}
  \centering
  \includegraphics[height=\linewidth,angle=270]{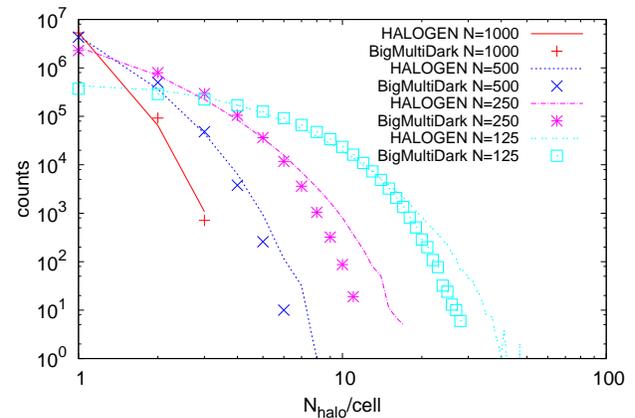}
  \caption{PDF of halo counts for both \halogen\ (lines) and \bigmd\ (points) catalogues from \bigmd.
  Several mesh numbers are used, as labelled by colors, and these correspond to the physical scales of
  $2.5\hMpc$, $5\hMpc$, $10\hMpc$ and $20\hMpc$ respectively.
  }
  \label{fig:PDF}
\end{figure}

Covering the \bigmd\ simulation with meshes of various (regular) sizes, we show in \Fig{fig:PDF} a histogram of the number of haloes per cell for both the 
\halogen\  and \bigmd\  catalogues; the cell size ranges from $2.5$ to $10$\hMpc. We find good agreement, especially at lower numbers of $N_{\rm halo}/$cell, 
where the contribution of non-linear scales is reduced. We note that the mesh used to calculate the PDF is not to be confused with the grid used by \halogen\ 
for the NGP density assignment.

\subsection{Power Spectrum}
\label{sec:pk}
\halogen\ has been designed to recover the 2PCF $\xi(r)$ of a provided halo catalogue. As the power spectrum $P(k)$ is its Fourier Transform,
it theoretically contains the same information. However, this information is distributed differently 
in the two functions and there is mode coupling when transforming from one to another: an error at a given scale in one of the magnitudes can propagate to
an error at all scales in the other.
So we expect to witness different strengths and weaknesses in $P(k)$.

\begin{figure}
  \centering
  \includegraphics[height=\linewidth,angle=270]{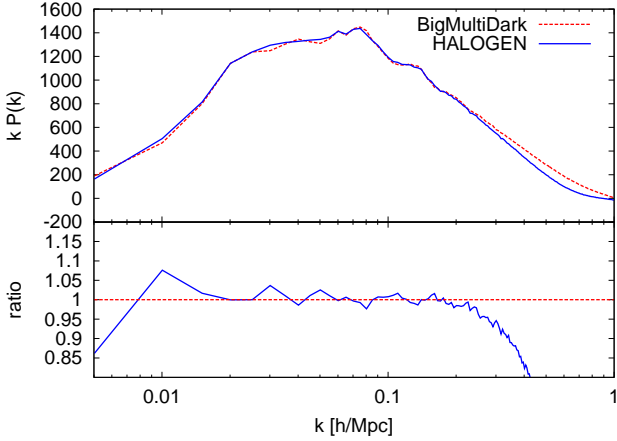}
  \caption{Power Spectrum P(k) of \halogen\ (blue line) and \fof\ (red line) for \bigmd.
The bottom panel shows their ratio. The Power Spectrum has been computed using a $N=1024^3$ mesh and corrected for shot noise as explained in \citet{jing05}.
  }
  \label{fig:Pk}
\end{figure}

In \Fig{fig:Pk} we compare the power spectrum of the \bigmd\ \fof\ catalogue to the corresponding \halogen\ realisation.  
We find agreement to 5\% across the scales $0.01h$Mpc$^{-1}<k<0.3 h$Mpc$^{-1}$, but note that smaller scales $k > 0.3h$Mpc$^{-1}$ ($r < 20\hMpc$) are underestimated. 
This underestimation arises from the smallest scales of the 2PCF, $r< l_{\rm cell}$, which integrate through higher scales in $P(k)$.
  
 \subsection{Correlation Function in Redshift Space}
 \label{sec:RSD}
Observed galaxies are not directly located in 3D space, 
 but in 2D-angular $(\theta, \phi)$ coordinates with redshift $z$ converted to a polar distance.
However, such distances are modified by galaxies' peculiar velocities -- velocity components that are not due to the Hubble expansion.
These modifications are encoded as Redshift Space Distortions (RSD), and we can begin to account for them by assigning correct velocities to haloes.

\begin{figure}
  \centering
  \includegraphics[height=\linewidth,angle=270]{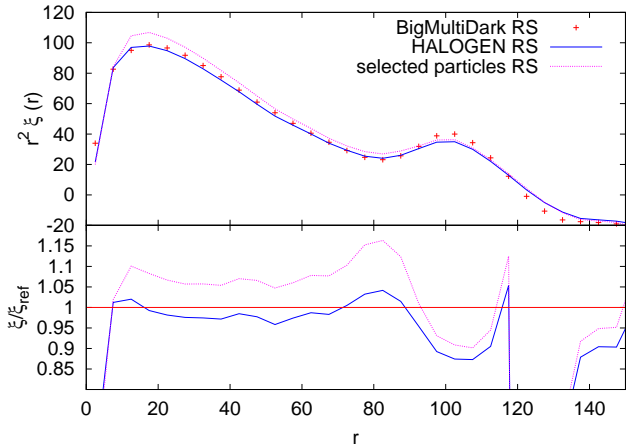}
  \caption{2PCF in redshift space (RS) for \fof\ (red points),
  and \halogen\ (blue line) of the \bigmd\ simulation. 
  We also include in magenta the results of our catalogue without applying the velocity bias 
  (i.e. $f_{\rm vel}=1$, 'selected particles') and find that a correct
  velocity bias is needed.}
  \label{fig:RSD}
\end{figure}

 Using the halo velocities, we can mimic this effect when calculating the 2PCF. We show the results of such an analysis in \Fig{fig:RSD}, in which the monopole
 of the 2PCF in redshift space is compared for the \halogen\ and \bigmd\ catalogues. 
To show the effect of our velocity transformation, we also include the 2PCF of the 'selected particles' in which the velocities were not transformed.
The normalisation and shape are significantly improved by the simple linear transformation (\Eq{eq:velbias}), and we find agreement to below 5\% per cent at intermediate scales.

\section{Conclusions}
We have presented a new method called \halogen\ for the construction of approximate halo catalogues. 
It consists of 4 major steps:

\begin{enumerate}
	\item create a distribution of particles in a cosmological volume using $2^{\rm nd}$-order Lagrangian Perturbation Theory 
	and distribute them in a grid of cell size $l_{\rm cell}$
	\item sample a theoretical halo mass function $n(>M)$ with a list of $N_{\rm h}$ halo masses $M$ and order them in descending mass.
	\item place the haloes at the position of particles with a probability dependent on the cell density and halo mass $P_{\rm cell} \propto \rho_{\rm cell} ^{\alpha(M)}$.
	We select random particles within cells, respecting the \textit{exclusion} criterion and conserving mass in cells (cf. \Sec{sec:halogen}).
	\item assign the velocity of the selected particle to the halo through a factor $\mathbf{v}_{\rm halo}=f_{\rm vel}(M)\cdot \mathbf{v}_{\rm part}$
\end{enumerate}

We noted the modularity of these steps and acknowledged alternatives for each of them.
The 2LPT in step (i) provides us with the correct large scale clustering at a low computational cost,
while step (ii) reconstructs the halo mass function.
The heart of \halogen\ is step (iii) where the mass dependent bias is modelled through the parameter $\alpha (M)$ 
that stochastically places more massive haloes in 
overdensities, recovering the correct 2-point correlation function as a function of mass. We also 
preclude haloes from overlapping to match the small-scale behaviour of the 2-point clustering.
In the last step (iv), we re-map particle velocities in order to obtain the correct halo velocity distribution. 

We studied how the parameters of the method -- $\alpha(M)$, $f_{\rm vel}(M)$ and $l_{\rm cell}$-- can be optimised and summarised the results in \Tab{tab:parameters}. 
Though \halogen\ needs a reference halo catalogue from an N-Body simulation to obtain $\alpha(M)$ and $f_{\rm vel}(M)$, once they have been optimised for a given setup,
\halogen\ can be used to generate
a multitude of halo catalogues, allowing the quantification of cosmic variance.

The halo mass function is recovered by construction --with some negligible sampling noise-- to the theoretical value.
The 2-point function at intermediate scales ($10\hMpc<r<50\hMpc$, where the bias is controlled by $\alpha(M)$) can be obtained in a \bigmd-like simulation at the $\sim2\%$ level 
and to the $15 \%$ level at BAO scales ($80\hMpc<r<110\hMpc$) (\Fig{fig:variance}). 
In redshift space, the error at intermediate scales rises to 
$\sim4\%$ and remains at $\sim15\%$ at large scales (\Fig{fig:RSD}). 
The clustering has a mass-dependence, for which the accuracy is controlled by the number of bins in the $\alpha(M)$ fit (\Fig{fig:MassBins}). 
The power spectrum can be recovered at the $5\%$ level in the range of scales $0.01{\rm Mpc}^{-1} h<k<0.3{\rm Mpc}^{-1} h$ (\Fig{fig:Pk}).
The halo PDF is accurately reproduced at low $N_{halo}/$cell, but overpredicts the high-$N_{\rm halo}$/cell tail where the contributions of non-linearities are higher
(\Fig{fig:PDF}).

We remark upon the adaptability of \halogen\ to different setups: \goliat\ and \bigmd\ have  different characteristics (see \Tab{tab:sims}) and \halogen\ can be used 
for both with little recalibration effort. This indicates that \halogen\ is not only capable of running on one specific boxsize, redshift or cosmology, which
makes it a powerful tool for exploring the statistics of varying cosmologies etc. 

We have also verified that changing the initial phases in \textsc{2LPTic} for \halogen\ leads to changes in the correlation function (due to cosmic variance) 
that follow the \nbody\ simulation.
This implies that doing so will yield robust estimates of cosmic variance, over potentially hundreds to thousands of realisations. 

We have demonstrated that \halogen\ is a powerful tool for modelling statistics of halo catalogues, and the effects cosmic variance on them. 
The most immediate application of \halogen\ is the generation of the many catalogues required to study the control of systematics and for 
computing covariance matrices for large galaxy surveys (e.g. DES, DESi, Euclid). However, it can conceivably be used for other applications involving the study of cosmic variance. 

Future work will involve improvements to the method, for instance by exploring sub-cell adjustments (i.e. alternatives to the random choice inside cells) or by 
changing one of the 4 stages of \halogen\ (e.g. what happens if we use 3LPT?, what is the best function for $G(\rc)$ in \Eq{eq:prob}?).
Furthermore, in the present study we have neglected substructure and refered to a possible extension using HOD models. We anticipate a fully integrated HOD layer to the method in future releases, which will enable a more direct comparison to observed data.


\section*{Acknowledgements}

SA and JGB acknowledge financial support from the Spanish MINECO under grant FPA2012-39684-C03-02 and Consolider-Ingenio ``Physics of the Accelerating Universe (PAU)" (CSD2007-00060).
They also acknowledge the support from the Spanish MINECO's ``Centro de Excelencia Severo Ochoa" Programme under Grant No. SEV-2012-0249.

SA is also supported by a PhD FPI-fellowship from the Universidad Aut\'onoma de Madrid. He also thanks the "Estancias Breves'' program from the UAM 
and the UWA Reasearch Collaboration Award 2014 that supported his stay in ICRAR, where this project was born.
He further thanks David Alonso for his advices at different stages of the project.

AK is supported by the {\it Ministerio de Econom\'ia y Competitividad} (MINECO) in Spain through grant AYA2012-31101 as well as the Consolider-Ingenio 2010 Programme of the {\it Spanish Ministerio de Ciencia e Innovaci\'on} (MICINN) under grant MultiDark CSD2009-00064. He also acknowledges support from the {\it Australian Research Council} (ARC) grants DP130100117 and DP140100198. He further thanks Dinosaur Jr. for the bug.

Part of this research was undertaken as part of the Survey Simulation 
Pipeline (SSimPL; {\small ssimpl-universe.tk}). The Centre for All-Sky 
Astrophysics (CAASTRO) is an Australian Research Council Centre of Excellence, 
funded by grant CE11E0090.

The work was supported by iVEC through the use of advanced computing resources
located at iVEC@Murdoch.

The MultiDark Database and the web application providing online access to it were constructed as part of the activities of the German Astrophysical Virtual Observatory as result of a collaboration between the Leibniz-Institute for Astrophysics Potsdam (AIP) and the Spanish MultiDark Consolider Project CSD2009-00064. The BigMD simulation suite have been performed in the Supermuc supercomputer at LRZ using time granted. The simulation and its FOF halo catalogue has been kindly made available to us courtesy Stefan Gottl\"ober, Anatoly Klypin, Francisco Prada, and Gustavo Yepes before its public release. We also acknowledge PRACE for awarding us access to resource Curie supercomputer based in France (project PA2259). Some computation were performed on HYDRA, the HPC-cluster of the IFT-UAM/CSIC.

This research has made use of NASA's Astrophysics Data System (ADS) and the arXiv preprint server.

\bibliography{mn-jour,refs}

\begin{thebibliography}{}

\bibitem[\protect\citeauthoryear{{Alonso}}{{Alonso}}{2012}]{CUTE}
{Alonso} D.,  2012, ArXiv e-prints 1210.1833

\bibitem[\protect\citeauthoryear{{Bond}, {Cole}, {Efstathiou} \&
  {Kaiser}}{{Bond} et~al.}{1991}]{EPS}
{Bond} J.~R.,  {Cole} S.,  {Efstathiou} G.,    {Kaiser} N.,  1991, \apj, 379,
  440

\bibitem[\protect\citeauthoryear{{Bouchet}, {Colombi}, {Hivon} \&
  {Juszkiewicz}}{{Bouchet} et~al.}{1995}]{bouchet}
{Bouchet} F.~R.,  {Colombi} S.,  {Hivon} E.,    {Juszkiewicz} R.,  1995, \aap,
  296, 575

\bibitem[\protect\citeauthoryear{{Castander}, {Ballester}, {Bauer},
  {Cardiel-Sas}, {Carretero}, {Casas}, {Castilla}, {Crocce}, {Delfino},
  {Eriksen} \& et al.}{{Castander} et~al.}{2012}]{PAU}
{Castander} F.~J.,  {Ballester} O.,  {Bauer} A.,  {Cardiel-Sas} L.,
  {Carretero} J.,  {Casas} R.,  {Castilla} J.,  {Crocce} M.,  {Delfino} M.,
  {Eriksen} M.,    et al. 2012, in Society of Photo-Optical Instrumentation
  Engineers (SPIE) Conference Series Vol.~8446 of Society of Photo-Optical
  Instrumentation Engineers (SPIE) Conference Series, {The PAU camera and the
  PAU survey at the William Herschel Telescope}.
p.~6

\bibitem[\protect\citeauthoryear{{Chuang}, {Kitaura}, {Prada}, {Zhao} \&
  {Yepes}}{{Chuang} et~al.}{2015}]{EZmocks}
{Chuang} C.-H.,  {Kitaura} F.-S.,  {Prada} F.,  {Zhao} C.,    {Yepes} G.,
  2015, \mnras, 446, 2621

\bibitem[\protect\citeauthoryear{{Chuang}, {Zhao}, {Prada}, {Munari}, {Avila},
  {Izard}, {Kitaura}, {Manera}, {Monaco}, {Murray}, {Knebe}, {Scoccola},
  {Yepes}, {Garcia-Bellido}, {Marin}, {Muller}, {Skibba}, {Crocce}, {Fosalba},
  {Gottlober}, {Klypin} \& {Power}}{{Chuang} et~al.}{2014}]{nifty}
{Chuang} C.-H.,  {Zhao} C.,  {Prada} F.,  {Munari} E.,  {Avila} S.,  {Izard}
  A.,  {Kitaura} F.-S.,  {Manera} M.,  {Monaco} P.,  {Murray} S.,  {Knebe} A.,
  {Scoccola} C.~G.,  {Yepes} G.,  {Garcia-Bellido} J.,  {Marin} F.~A.,
  {Muller} V.,  {Skibba} R.,  {Crocce} M.,  {Fosalba} P.,  {Gottlober} S.,
  {Klypin} A.~A.,  {Power} C.,  {Tao} C.,    {Turchaninov} V.,  2014, ArXiv
  e-prints 1412.5228

\bibitem[\protect\citeauthoryear{{Coles} \& {Jones}}{{Coles} \&
  {Jones}}{1991}]{lognormal}
{Coles} P.,  {Jones} B.,  1991, \mnras, 248, 1

\bibitem[\protect\citeauthoryear{{Col{\'{\i}}n}, {Klypin} \&
  {Kravtsov}}{{Col{\'{\i}}n} et~al.}{2000}]{colin2000}
{Col{\'{\i}}n} P.,  {Klypin} A.~A.,    {Kravtsov} A.~V.,  2000, \apj, 539, 561

\bibitem[\protect\citeauthoryear{{Davis}, {Efstathiou}, {Frenk} \&
  {White}}{{Davis} et~al.}{1985}]{Davis85}
{Davis} M.,  {Efstathiou} G.,  {Frenk} C.~S.,    {White} S.~D.~M.,  1985, \apj,
  292, 371

\bibitem[\protect\citeauthoryear{{Dawson}, {Schlegel}, {Ahn}, {Anderson},
  {Aubourg}, {Bailey}, {Barkhouser}, {Bautista}, {Beifiori} \& et al.}{{Dawson}
  et~al.}{2013}]{BOSS}
{Dawson} K.~S.,  {Schlegel} D.~J.,  {Ahn} C.~P.,  {Anderson} S.~F.,  {Aubourg}
  {\'E}.,  {Bailey} S.,  {Barkhouser} R.~H.,  {Bautista} J.~E.,  {Beifiori} A.,
     et al. 2013, \aj, 145, 10

\bibitem[\protect\citeauthoryear{{Frieman} \& {Dark Energy Survey
  Collaboration}}{{Frieman} \& {Dark Energy Survey Collaboration}}{2013}]{DES}
{Frieman} J.,  {Dark Energy Survey Collaboration} 2013, in American
  Astronomical Society Meeting Abstracts 221 Vol.~221 of American Astronomical
  Society Meeting Abstracts, {The Dark Energy Survey: Overview}.
p. 335.01

\bibitem[\protect\citeauthoryear{{Fry}}{{Fry}}{1985}]{CIC-2}
{Fry} J.~N.,  1985, \apj, 289, 10

\bibitem[\protect\citeauthoryear{{Hockney} \& {Eastwood}}{{Hockney} \&
  {Eastwood}}{1988}]{Hockney:1988}
{Hockney} R.~W.,  {Eastwood} J.~W.,  1988, {Computer simulation using
  particles}

\bibitem[\protect\citeauthoryear{{Jing}}{{Jing}}{2005}]{jing05}
{Jing} Y.~P.,  2005, \apj, 620, 559

\bibitem[\protect\citeauthoryear{{Kitaura}, {Yepes} \& {Prada}}{{Kitaura}
  et~al.}{2014}]{patchy}
{Kitaura} F.-S.,  {Yepes} G.,    {Prada} F.,  2014, \mnras, 439, L21

\bibitem[\protect\citeauthoryear{{Klypin}, {Yepes}, {Gottlober}, {Prada} \&
  {Hess}}{{Klypin} et~al.}{2014}]{bigmd}
{Klypin} A.,  {Yepes} G.,  {Gottlober} S.,  {Prada} F.,    {Hess} S.,  2014,
  ArXiv e-prints 1411.4001

\bibitem[\protect\citeauthoryear{{Knebe}, {Knollmann}, {Muldrew}, {Pearce} \&
  {et al.}}{{Knebe} et~al.}{2011}]{Knebe11}
{Knebe} A.,  {Knollmann} S.~R.,  {Muldrew} S.~I.,  {Pearce} F.~R.,    {et al.}
  2011, \mnras, 415, 2293

\bibitem[\protect\citeauthoryear{{Knebe}, {Pearce}, {Lux}, {Ascasibar},
  {Behroozi}, {Casado}, {Moran}, {Diemand} \& {et al.}}{{Knebe}
  et~al.}{2013}]{Knebe:2013}
{Knebe} A.,  {Pearce} F.~R.,  {Lux} H.,  {Ascasibar} Y.,  {Behroozi} P.,
  {Casado} J.,  {Moran} C.~C.,  {Diemand} J.,    {et al.} 2013, \mnras, 435,
  1618

\bibitem[\protect\citeauthoryear{{Knollmann} \& {Knebe}}{{Knollmann} \&
  {Knebe}}{2009}]{ahf}
{Knollmann} S.~R.,  {Knebe} A.,  2009, \apjs, 182, 608

\bibitem[\protect\citeauthoryear{{Laureijs}, {Amiaux}, {Arduini},
  {Augu{\`e}res}, {Brinchmann}, {Cole}, {Cropper}, {Dabin}, {Duvet}, {Ealet} \&
  et al.}{{Laureijs} et~al.}{2011}]{Euclid}
{Laureijs} R.,  {Amiaux} J.,  {Arduini} S.,  {Augu{\`e}res} J.~.,  {Brinchmann}
  J.,  {Cole} R.,  {Cropper} M.,  {Dabin} C.,  {Duvet} L.,  {Ealet} A.,    et
  al. 2011, ArXiv e-prints 1110.3193

\bibitem[\protect\citeauthoryear{{Levi}, {Bebek}, {Beers}, {Blum}, {Cahn},
  {Eisenstein}, {Flaugher}, {Honscheid}, {Kron}, {Lahav}, {McDonald}, {Roe},
  {Schlegel} \& {representing the DESI collaboration}}{{Levi}
  et~al.}{2013}]{DESi}
{Levi} M.,  {Bebek} C.,  {Beers} T.,  {Blum} R.,  {Cahn} R.,  {Eisenstein} D.,
  {Flaugher} B.,  {Honscheid} K.,  {Kron} R.,  {Lahav} O.,  {McDonald} P.,
  {Roe} N.,  {Schlegel} D.,    {representing the DESI collaboration} 2013,
  ArXiv e-prints 1308.0847

\bibitem[\protect\citeauthoryear{{Manera}, {Scoccimarro} \&
  {Percival}}{{Manera} et~al.}{2013}]{PTHalos_MM}
{Manera} M.,  {Scoccimarro} R.,    {Percival} W.~J. e.~a.,  2013, \mnras, 428,
  1036

\bibitem[\protect\citeauthoryear{{Monaco}, {Sefusatti}, {Borgani}, {Crocce},
  {Fosalba}, {Sheth} \& {Theuns}}{{Monaco} et~al.}{2013}]{pinocchio_2}
{Monaco} P.,  {Sefusatti} E.,  {Borgani} S.,  {Crocce} M.,  {Fosalba} P.,
  {Sheth} R.~K.,    {Theuns} T.,  2013, \mnras, 433, 2389

\bibitem[\protect\citeauthoryear{{Monaco}, {Theuns} \& {Taffoni}}{{Monaco}
  et~al.}{2002}]{pinocchio_1}
{Monaco} P.,  {Theuns} T.,    {Taffoni} G.,  2002, \mnras, 331, 587

\bibitem[\protect\citeauthoryear{{Moutarde}, {Alimi}, {Bouchet}, {Pellat} \&
  {Ramani}}{{Moutarde} et~al.}{1991}]{2lpt_0}
{Moutarde} F.,  {Alimi} J.-M.,  {Bouchet} F.~R.,  {Pellat} R.,    {Ramani} A.,
  1991, \apj, 382, 377

\bibitem[\protect\citeauthoryear{{Murray}, {Power} \& {Robotham}}{{Murray}
  et~al.}{2013}]{hmfcalc}
{Murray} S.~G.,  {Power} C.,    {Robotham} A.~S.~G.,  2013, Astronomy and
  Computing, 3, 23

\bibitem[\protect\citeauthoryear{{Neyrinck}}{{Neyrinck}}{2013}]{shell-crossing}
{Neyrinck} M.~C.,  2013, \mnras, 428, 141

\bibitem[\protect\citeauthoryear{{Peebles}}{{Peebles}}{1980}]{CIC}
{Peebles} P.~J.~E.,  1980, in {Ehlers} J.,  {Perry} J.~J.,   {Walker} M.,  eds,
  Ninth Texas Symposium on Relativistic Astrophysics Vol.~336 of Annals of the
  New York Academy of Sciences, {Statistics of the distribution of galaxies}.
pp 161--171

\bibitem[\protect\citeauthoryear{{Planck Collaboration}}{{Planck
  Collaboration}}{2014}]{planck}
{Planck Collaboration} 2014, \aap, 571, A16

\bibitem[\protect\citeauthoryear{{Press} \& {Schechter}}{{Press} \&
  {Schechter}}{1974}]{Press1974}
{Press} W.~H.,  {Schechter} P.,  1974, \apj, 187, 425

\bibitem[\protect\citeauthoryear{{Sahni} \& {Shandarin}}{{Sahni} \&
  {Shandarin}}{1996}]{shell-crossing-2}
{Sahni} V.,  {Shandarin} S.,  1996, \mnras, 282, 641

\bibitem[\protect\citeauthoryear{{Saslaw}}{{Saslaw}}{2000}]{CIC-3}
{Saslaw} W.~C.,  2000, {The Distribution of the Galaxies}

\bibitem[\protect\citeauthoryear{{Scoccimarro}}{{Scoccimarro}}{1998}]{2lptcode}
{Scoccimarro} R.,  1998, \mnras, 299, 1097

\bibitem[\protect\citeauthoryear{{Scoccimarro} \& {Sheth}}{{Scoccimarro} \&
  {Sheth}}{2002}]{PTHalos}
{Scoccimarro} R.,  {Sheth} R.~K.,  2002, \mnras, 329, 629

\bibitem[\protect\citeauthoryear{{Skibba} \& {Sheth}}{{Skibba} \&
  {Sheth}}{2009}]{Skibba2009}
{Skibba} R.~A.,  {Sheth} R.~K.,  2009, \mnras, 392, 1080

\bibitem[\protect\citeauthoryear{{Springel}}{{Springel}}{2005}]{gadget}
{Springel} V.,  2005, \mnras, 364, 1105

\bibitem[\protect\citeauthoryear{{Tassev}, {Zaldarriaga} \&
  {Eisenstein}}{{Tassev} et~al.}{2013}]{cola}
{Tassev} S.,  {Zaldarriaga} M.,    {Eisenstein} D.~J.,  2013, Journal of
  Cosmology and Astroparticle Physics, 6, 36

\bibitem[\protect\citeauthoryear{{Tinker}, {Kravtsov}, {Klypin}, {Abazajian},
  {Warren}, {Yepes}, {Gottl{\"o}ber} \& {Holz}}{{Tinker}
  et~al.}{2008}]{Tinker08}
{Tinker} J.,  {Kravtsov} A.~V.,  {Klypin} A.,  {Abazajian} K.,  {Warren} M.,
  {Yepes} G.,  {Gottl{\"o}ber} S.,    {Holz} D.~E.,  2008, \apj, 688, 709

\bibitem[\protect\citeauthoryear{{Tinker}, {Weinberg}, {Zheng} \&
  {Zehavi}}{{Tinker} et~al.}{2005}]{Tinker05}
{Tinker} J.~L.,  {Weinberg} D.~H.,  {Zheng} Z.,    {Zehavi} I.,  2005, \apj,
  631, 41

\bibitem[\protect\citeauthoryear{{Watson}, {Iliev}, {D'Aloisio}, {Knebe},
  {Shapiro} \& {Yepes}}{{Watson} et~al.}{2013}]{watson}
{Watson} W.~A.,  {Iliev} I.~T.,  {D'Aloisio} A.,  {Knebe} A.,  {Shapiro} P.~R.,
     {Yepes} G.,  2013, \mnras, 433, 1230

\bibitem[\protect\citeauthoryear{{White}, {Tinker} \& {McBride}}{{White}
  et~al.}{2014}]{qpm}
{White} M.,  {Tinker} J.~L.,    {McBride} C.~K.,  2014, \mnras, 437, 2594

\end{thebibliography}
\bibliographystyle{mn2e} \label{sec:Bibliography}

 \label{lastpage}
\end{document}